\documentclass[twocolumn,final]{revtex4-1}
\usepackage{graphicx}% Include figure files
\usepackage{setspace}
\usepackage{caption}
\usepackage{subcaption}
\usepackage{color}
\usepackage{dcolumn}% Align table columns on decimal point
\usepackage{bm}% bold math
\usepackage{amssymb,amsfonts}
\usepackage{tabularx}
\usepackage{comment}
\usepackage[mathscr]{euscript}
\usepackage{makecell}
\usepackage{wrapfig}
\usepackage{graphicx}
\usepackage{bm}
\usepackage{amsmath}
\usepackage{amssymb}
\usepackage{caption}
\usepackage{hyperref}
\usepackage{tikz}
\usepackage{booktabs}
\usetikzlibrary{arrows.meta, positioning}
\usepackage{xcolor}
\usepackage{xspace}
\usepackage{array}
\usepackage{booktabs}

\usetikzlibrary{arrows.meta}
\usetikzlibrary{positioning}
\usetikzlibrary{calc}
\usetikzlibrary{shapes.multipart}
\usetikzlibrary{shapes.geometric}
\usetikzlibrary{fit}
\usetikzlibrary{decorations.pathmorphing}

\begin{document}

\newcommand{\la}{\left\langle}
\newcommand{\ra}{\right\rangle}
\newcommand{\lu}{\left}
\newcommand{\ru}{\right}
\newcommand{\be}{\begin{equation}}
\newcommand{\ee}{\end{equation}}
\newcommand{\bea}{\begin{eqnarray}}
\newcommand{\eea}{\end{eqnarray}}
\newcommand{\dst}{\displaystyle}
\newcommand{\ovn}{\overline}
\newcommand{\vre}{\varepsilon}
\newcommand{\eps}{\epsilon}
\newcommand{\vph}{\varphi}
\newcommand{\tf}{\textbf}
\newcommand{\rea}{\text{Re}}
\newcommand{\ima}{\text{Im}}
\newcommand{\Ai}{\text{Ai}}
\newcommand{\Bi}{\text{Bi}}
\newcommand{\te}{\text}
\newcommand{\ti}{\tilde}
\newcommand{\pT}{\partial}
\newcommand{\pa}{\parallel}
\newcommand{\x}{\ti x}
\newcommand{\y}{\ti y}
\newcommand{\z}{\ti z}
\newcommand{\ts}{\ti t}
\newcommand{\mf}{\te {\bf m}}
\newcommand{\xfrac}[2]{\genfrac{}{}{0pt}{0}{#1}{#2}}
\newcommand{\newfrac}[2]{\genfrac{}{}{}{0}{#1}{#2}}
\newcommand{\widearc}{\wideparen}
\newcommand{\F}{\mathcal{F}}
\newcommand{\GKFieldFlow}{\textsc{GKFieldFlow}\xspace}
\newcommand{\GKFieldFlowNet}{\textsc{GK-FieldFlow-Net}\xspace}
\newcommand{\FieldFlowNet}{\textsc{FieldFlow-Net}\xspace}

\bibliographystyle{iopart-num}

\title{\textbf{GKFieldFlow: A Spatio-Temporal Neural Surrogate for Nonlinear Gyrokinetic Turbulence}\\
\large \textit{A 3D U-Net + TCN Architecture for Joint Field \& Transport Prediction}}

\author{Arash Ashourvan}
\affiliation{Independent Researcher, San Diego, CA, USA}

%\ead{submissions@iop.org}
\vspace{10pt}
\begin{abstract}
We present \GKFieldFlow, a novel three-dimensional autoregressive deep learning 
surrogate model for nonlinear gyrokinetic turbulence. Based on the 
architecture \FieldFlowNet{}, this model combines a multi–resolution 3D U–Net 
encoder-decoder that operates on evolving plasma potential fields. A dilated temporal convolutional network 
(TCN) learns the nonlinear time evolution of latent turbulence features. \GKFieldFlow simultaneously (i) predicts ion and electron energy fluxes, 
and particle flux directly from CGYRO turbulence, and (ii) predicts
future potential fields autoregressively with desired spatial resolution. This enables the model to replicate both instantaneous transport and the 
underlying spatio–temporal dynamics that generate it.

The architecture is physics‐informed in its design: 3D convolutions preserve 
the anisotropic geometry and phase structure of gyrokinetic fluctuations, while 
dilated temporal convolutions capture multiscale dynamical couplings such as turbulence and zonal‐flow interactions, turbulence decorrelation, and intermittent bursty transport.  We provide a complete technical description of the data structure, 
model components, and rationale behind each architectural choice.

The model achieves high accuracy across all three transport channels, 
with multi–horizon inference maintaining robustness. Autoregressive field rollouts preserve the spectral content, phase coherence, and energy distribution of the CGYRO 
nonlinear state with strong fidelity, and flux predictions remain consistent 
with CGYRO within a small fractional error. This work presents \GKFieldFlow{} as a data‐driven reduced model that can jointly learn turbulence dynamics and transport.

\end{abstract}

\maketitle

\section{Introduction}

Understanding and predicting turbulent transport remains a central challenge in
magnetic-confinement fusion research. Small-scale microturbulence regulates
cross-field heat and particle fluxes in tokamak plasmas and strongly influences
macroscopic confinement performance. As a result, the ability to model,
forecast, and control turbulent transport is essential for improving predictive
capabilities and advancing toward reliable, reactor-relevant plasma operation.
\cite{horton1999,tang2011,kotschenreuther2019}.

Nonlinear gyrokinetic (GK) simulations provide high-fidelity predictions of microturbulence and transport in magnetically confined fusion plasmas.
However, despite their accuracy, these simulations remain computationally demanding.
A single ion-scale CGYRO simulation typically requires hundreds of GPU‐hours to reach a
fully saturated turbulent state, and even more when scans over equilibrium profiles,
collisionality, $\beta$, or shaping parameters are required.  
The challenge becomes dramatically more severe for plasmas exhibiting
\emph{multiscale turbulence}, where long-wavelength ion-scale modes interact with
short-wavelength electron‐scale modes. Fully resolved multi-scale simulations require
simultaneous retention of disparate spatial and temporal scales, increasing
computational cost by two to three orders of magnitude
\cite{howard2016,mandell2018,maeyama2021}.  
As a result, routine predictive modeling based on nonlinear GK turbulence remains out of reach for most experimental workflows, scenario development studies, and real-time
plasma control applications.
Instead, transport codes rely almost exclusively on quasilinear (QL) reduced models,
such as TGLF~\cite{tglf2007} and QuaLiKiz~\cite{qualikiz2009} for predictive modeling, which use a simplified
approximation of the nonlinear saturation spectrum to compute transport fluxes at
orders of magnitude lower computational cost.  
While QL models have demonstrated impressive predictive capability in many core-plasma
regimes, their accuracy is known to degrade in strongly nonlinear or electromagnetic
edge conditions, including pedestal regions with steep gradients, strong $E \times B$
shear, microtearing turbulence, and regimes where multiscale or nonlocal effects are
important \cite{ashourvan2024,groebner2018,shi2019}.  
These limitations highlight a growing need for new turbulence surrogates that retain
the fidelity of nonlinear GK modeling while achieving significantly faster inference.

The emergence of modern machine-learning(ML) provides a promising path forward.
The fusion community now possesses an extensive archive of both linear and nonlinear
gyrokinetic simulations spanning decades of theory and experiment. Leveraging these
databases to train high‐dimensional neural surrogate models has the potential to
dramatically accelerate turbulence inference, enable rapid profile prediction, and
support integrated modeling workflows for devices such as ITER and future pilot plants.
However, teaching a neural network to emulate the rich, multiscale, and anisotropic
structure of gyrokinetic turbulence remains an open challenge.

\begin{figure*}[t]
\centering
\includegraphics[width=0.8\textwidth]{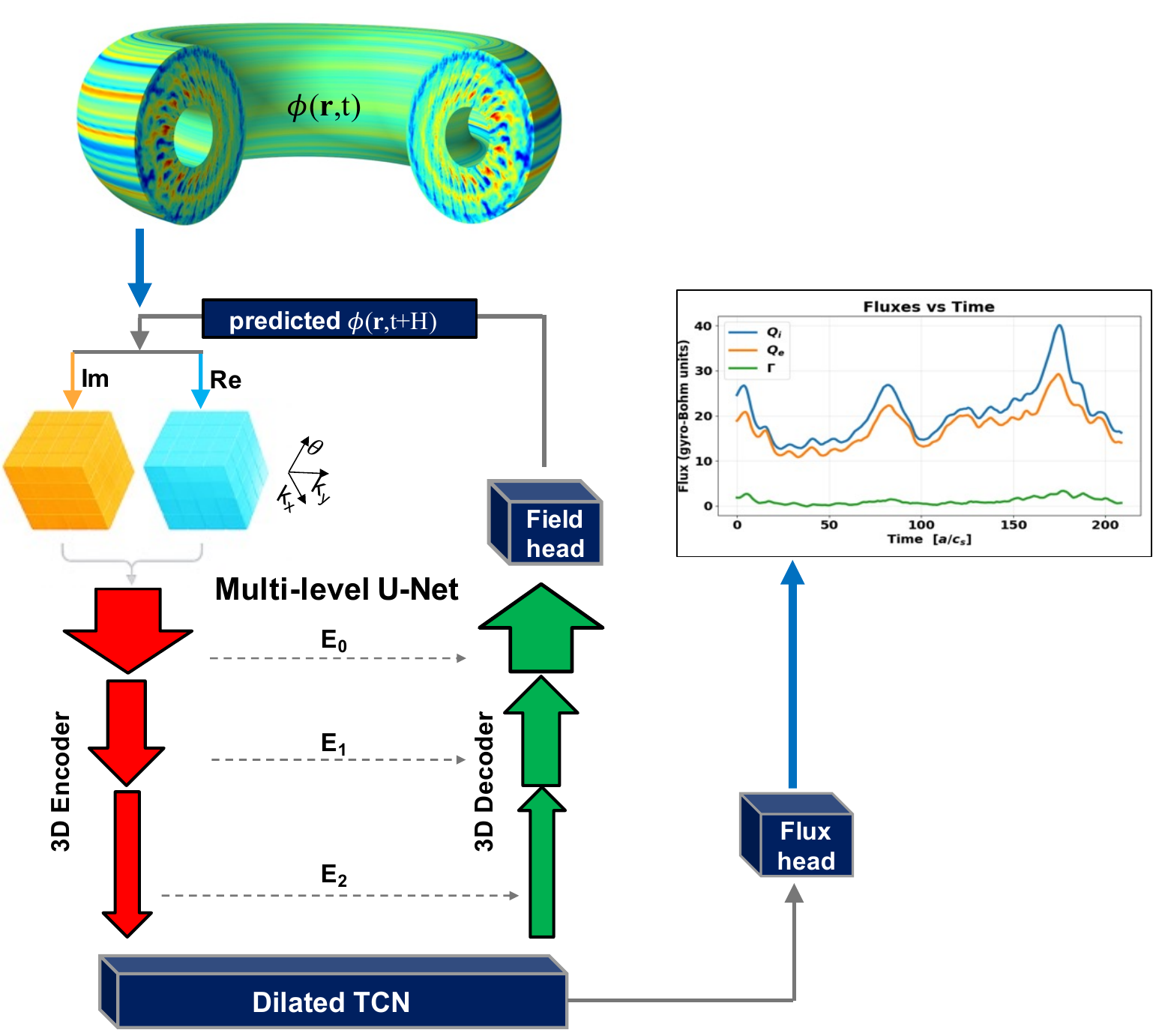}
\caption{Schematic overview of the \GKFieldFlow model and the autoregressive  field potential calculation loop. Turbulence field potentials $\phi(\mathbf{r},t)$ and fluxes ($Q_i,Q_e,\Gamma$) used for training the neural-net are pre-calculated by the CGYRO nonlinear gyrokinetic code. The \FieldFlowNet architecture is designed as follows: (i) a multi-level 3D U-Net encoder extracts spatial turbulence features from a short history of complex fields (Re/Im channels) (ii) a dilated temporal convolutional network (TCN) models temporal dependencies across snapshots to form a final latent representation, which feeds two heads: a field head driving a 3D decoder to predict $\widehat{\Phi}(\mathbf{r},t+H)$ and a flux head predicting $(Q_i,Q_e,\Gamma)$ over forecast horizons. Exact layer dimensions and hyperparameters are omitted for clarity. }
\label{fig:overview}
\end{figure*}

In this work, we present \GKFieldFlow, a 3D spatio–temporal neural surrogate that
both evolves gyrokinetic turbulent structure and predicts the associated
transport, trained directly on nonlinear CGYRO field data. The model is realized
through a dedicated architecture that we refer to as \FieldFlowNet: a 3D
convolutional U-Net encoder--decoder coupled to a dilated temporal
convolutional network (TCN). Together, these components form a two-head
autoregressive surrogate that reconstructs future fields and forecasts transport  fluxes from fully nonlinear gyrokinetic simulations.

A Schematic overview of  \GKFieldFlow is shown in Fig.~\ref{fig:overview}. The model ingests the complex potential field $\phi(\mathbf{r},t)$ (here electrostatic), over a short temporal context window $T_c$, encodes the 3D spatial structure, and evolves a latent representation with a TCN. The input field is spectral in the perpendicular plane to the magnetic field $(k_x,k_y)$ and finite-difference in $\theta$, the poloidal angle which represents the variation along the magnetic field. Two output heads then (i) reconstruct the future $\phi$ field on the full $(k_x,k_y,\theta)$ grid and (ii) infer the corresponding ion and electron heat fluxes and particle flux, all in physical units.  We further introduce an internal rollout loss that exposes the model during training to multi-step prediction, and we use ensemble rollouts to quantify model fidelity over hundreds of time steps. 

Because the network lacks velocity-space information, species-dependent kinetic moments, 
the governing GK–Maxwell equations, and is sparsified in time and spatial information the evolution of fields is non-Markovian from the surrogate’s
perspective. A key result is the identification of an \emph{optimal memory time} $T^*_c$ that minimizes long-horizon RMSE of both fluxes and fields, which we show is significantly shorter than the turbulence autocorrelation time of the underlying CGYRO simulation.  This places our work at the intersection of ML-based surrogate modelling, gyrokinetic turbulence, and extreme-event prediction: we demonstrate that a relatively short, learned latent memory can sustain accurate long-rollout predictions of both transport and turbulence, suggesting new ways to build fast yet physics-aware surrogates for nonlinear gyrokinetic dynamics.

\begin{table*}[t]
\caption{Comparison of representative ML surrogates for plasma turbulence.
\GKFieldFlow targets nonlinear gyrokinetic turbulence with joint field and flux
forecasting over multiple horizons.}
\label{tab:surrogate_comparison}

\begin{ruledtabular}
\begin{tabular}{lll}
Model & Regime & Architecture \\ \hline

GAIT (Clavier 2025) &
Reduced HW model &
\makecell[l]{VAE + recurrent \\ generative decoder} \\

GyroSwin / CNN--LSTM &
\makecell[l]{GK/fluid \\ snapshots} &
\makecell[l]{2D CNN + recurrent \\ (LSTM/transformer)} \\

TGLF--SiNN &
\makecell[l]{Reduced GK \\ transport closure} &
\makecell[l]{NN regression \\ (no field evolution)} \\

Fluid U-Net surrogate baselines &
\makecell[l]{Non-plasma PDE / fluid turbulence \\ (no gyrokinetics)} &
\makecell[l]{Encoder--decoder U-Net for field regression; \\ single-step forecasts with no memory; \\ used in fluid/PDE surrogates \cite{ma2021physicscnn,bao2022physicsguided}}
\\

\hline
\textbf{\GKFieldFlow (this work)} &
\makecell[l]{Nonlinear GK \\ (CGYRO)} &
\makecell[l]{3D CNN encoder + \\TCN, joint \\ field+flux forecast} \\
%\multicolumn{3}{c}{\textit{Only model listed with stable multi-step rollouts and learned memory scale}} \\

\end{tabular}
\end{ruledtabular}
\end{table*}

\par 

\subsection{Machine-learning surrogates for plasma turbulence and gyrokinetic modeling}
\label{sec:ml_turbulence_gk}
Recent machine–learning surrogates for plasma turbulence span a range of
approaches depending on the retained physics and target observables. Generative
models such as GAIT \cite{clavier2025gait} demonstrate that latent autoencoders
and recurrent decoders can reproduce long sequences of turbulence in reduced
fluid models (e.g., Hasegawa–Wakatani), while CNN–LSTM and transformer
architectures have been applied to short–horizon field prediction from
gyrokinetic snapshots \cite{Boffi_2023}. Neural transport
regressors, including TGLF–SiNN~\cite{TGLF-SiNN} and related closure models
\cite{plass}, provide fast profile-level flux
estimates but do not evolve fields or maintain multi-step dynamical
consistency.

Outside the plasma community, encoder–decoder U-Nets are commonly used as
single-frame field regressors in fluid and PDE surrogate modeling
\cite{ma2021physicscnn,bao2022physicsguided}, but these architectures do not
include temporal memory, autoregressive stability, or joint transport
prediction. This work extends beyond those non-temporal U-Net baselines by
introducing a 3D spatio–temporal surrogate for nonlinear gyrokinetic turbulence,
in which a latent-memory model is trained directly on nonlinear CGYRO
data to jointly predict future fields and transport across multiple horizons.

A structured comparison is given in
Table~\ref{tab:surrogate_comparison} which summarizes these categories as they relate
to the present work. Prior surrogates typically (i) operate on reduced turbulence
models rather than first-principles gyrokinetics, (ii) generate fields without
predicting the associated transport, or (iii) regress transport without retaining
the underlying spatio–temporal structure. 
In contrast to prior ML surrogate, \GKFieldFlow jointly predicts future fields and fluxes by employing a 3D spatio–temporal neural architecture. This enables long-horizon, high-fidelity, autoregressive forecasting of nonlinear gyrokinetic turbulence with an effective computational speedup of \textbf{$\sim\mathcal{O}(10^2)\times - \mathcal{O}(10^3)\times$} in GPU resource usage relative to CGYRO for matched physical intervals for the representative DIII-D L-mode test case presented here.

\begin{comment}
While the proposed surrogate achieves low prediction error and stable autonomous rollouts, the primary purpose of the architectural scans presented in Figs.~\ref{fig:LayersvR} and~\ref{fig:Tc_scan} is to validate the functional role of the model components. We show that for the physics-based reduced cases of interest, Increasing the depth of the spatial UNet encoder systematically improves flux prediction accuracy, confirming that hierarchical spatial feature extraction is essential for capturing the multiscale structure of gyrokinetic turbulence. Independently, increasing the temporal context length $T_c$ leads to a clear reduction in error over a finite range, demonstrating that the temporal convolutional network (TCN) successfully exploits short-range temporal correlations across successive field snapshots. Together, these results show that the model is not acting as a purely instantaneous field-to-flux mapper: spatial multiscale information and temporal context both contribute meaningfully to predictive performance.
\end{comment}

The remainder of this article is organized as follows.
In Section~\ref{arch}, we introduce the \GKFieldFlow{} model and outline the underlying \FieldFlowNet{} architecture, including the data structure, the 3D spatial encoder–decoder, and the temporal convolutional network (TCN) used for latent evolution.
Section~\ref{crop} describes the reduction of the radial dimension using Lorentzian spectral widths.
In Section~\ref{depth}, we investigate the multilevel structure of the U-Net backbone and assess the impact of depth on performance.
Section~\ref{scan_tc} presents a scan over temporal context size $T_c$ and identifies an optimal window for TCN-based forecasting.
In Section~\ref{longh}, we evaluate long-horizon rollout stability and show that \GKFieldFlow{} maintains phase coherence, modal structure, and flux prediction accuracy over several turbulence autocorrelation times.
Finally, Section~\ref{summ} summarizes the results and discusses future developments.

\section{\label{arch}\FieldFlowNet architecture: a 3D U-Net encoder–decoder with temporal TCN dynamics}
\subsection{Architectural Lineage: 3D Convolutions, U-Net, and TCNs }
This subsection provides brief background on the core architectural elements—3D convolutional encoders, U-Net style spatial decoders, and causal temporal convolutional networks—and their relevance to spatio-temporal surrogate modeling in plasma turbulence.

Convolutional neural networks extended to volumetric data (3D CNNs) \cite{ji20123dcnn} have further 
allowed such architectures to operate natively on three-dimensional fields, making them especially 
well suited for plasma physics where gyrokinetic and reduced models produce inherently 
three-dimensional (or higher dimensional) structures. In particular, 3D convolutions provide 
translation-equivariant filtering in the triply periodic directions and can efficiently capture the 
local spatial coherence of turbulent fluctuations. Recent work has used 3D CNNs for 
gyrokinetic diagnostics, blob tracking, and volumetric reconstruction.\par

Neural-network architectures based on the U-Net \cite{ronneberger2015unet} have become the standard 
for learning structured mappings in high-dimensional spatial domains. Originally developed for 
biomedical image segmentation, the U-Net architecture introduced the concept of a hierarchical 
encoder–decoder structure augmented by skip connections between corresponding resolution levels. 
This design dramatically improves gradient flow while allowing high-resolution spatial information 
to bypass the deep bottleneck. Since then, U-Net variants have been used in a wide range of 
scientific applications, including turbulence modeling, climate prediction, volumetric image 
reconstruction, and multi-physics simulations.

To model temporal dynamics in a physically relevant manner, we employ a temporal convolutional 
network (TCN) \cite{lea2017temporal, bai2018tcn}, a causal architecture whose dilated 
one-dimensional convolutions provide a large receptive field without resorting to recurrent units. 
TCNs have been successfully used in video processing, climate time-series, and physics-informed 
sequence modeling, and they inherently preserve causality when predicting the next state given a 
window of past states. Importantly, dilated convolutions enable the network to learn multi-scale 
temporal structure, which is expected in gyrokinetic turbulence where decorrelation times, 
autocorrelation lengths, and zonal-flow oscillations coexist across widely separated scales.

\subsection{Overview of the autoregressive \GKFieldFlow model}

The model introduced here, which we refer to as the 
\textit{\GKFieldFlow U-Net–TCN autoregressive surrogate}, synthesizes these developments 
into a unified 3D spatio-temporal architecture designed specifically for the CGYRO 
gyrokinetic code. The goal is twofold:

\begin{enumerate}
    \item Predict the nonlinear evolution of the electrostatic potential 
    $\Phi(\textbf{r}, t)$ across an arbitrary forecast horizon 
    $H$, yielding $\Phi(\textbf{r},t+H)$; and
    \item Infer the corresponding ion heat flux ($Q_i$), electron heat flux ($Q_e$),
    and particle flux ($\Gamma$) directly from the latent temporal representation.
\end{enumerate}

The architecture of \GKFieldFlow is intentionally designed to reflect two key properties of gyrokinetic turbulence: strong multiscale spatial structure and finite, but limited, temporal dependence.
From now on we refer to the electrostatic field potentials used in \GKFieldFlow as $\Phi$ (in contrast to $\phi$) to distinguish them from CGYRO calculations. Unlike prior deep surrogate models trained on static snapshots or instantaneous fluxes, our method 
learns a \textit{temporal dynamical law} in latent space. This represents the first integration of:
(i) a 3D convolutional encoder operating directly on 
$(k_x,k_y,\theta)$ volumetric potential fields;
(ii) a U-Net style decoder reconstructing full-resolution future fields; and
(iii) a dilated TCN evolving the sequence of latent representations over time. Detailed specifications of encoder, decoder, and TCN layer configurations are provided in Appendix~\ref{app:arch}.

The architecture is thus autoregressive: inferences of $\Phi(\textbf{r},t+H)$ may be recursively fed back into 
the model to generate long-horizon trajectory rollouts, enabling reduced-cost prediction of turbulent 
dynamics.
\subsection{CGYRO Data and Physical Inputs}

\subsubsection{Electrostatic Potential Fields}

CGYRO outputs the 3D time evolving electrostatic potential as spectral in the radial and binormal direction and finite difference mesh in the poloidal direction  $\phi(\mathbf{r},t) \rightarrow \phi(k_x,k_y,\theta;t)$. 
At each saved time slice $t_k$, the model receives:

\begin{itemize}
    \item $\Re(\phi)$ --- real component,
    \item $\Im(\phi)$ --- imaginary component.
\end{itemize}

These two components are treated as distinct input channels, giving the input 
tensor shape:
\begin{equation}
    \phi_{\rm input}
    \in
    \mathbb{R}^{T \times 2 \times R \times \Theta \times N_{y}}.
\end{equation}
where $T$ is the number of time slices, the factor of $2$ corresponds to the real
and imaginary components of the complex potential,
$R$ is the number of retained radial Fourier (or spectral) modes, $\Theta$ is the number
of poloidal grid points in the $\theta$ direction, and $N_{y}$ is the number of
binormal (toroidal) Fourier modes.   
The decision to preserve both components as channels reflects the physical 
meaning of the quadrature components of the complex mode amplitude.  
Phase information plays a central role in determining nonlinear coupling, 
toroidal mode rotation, and the structure of turbulence eddies.
The input electrostatic potential $\Phi$ is supplied to the network in the form of a five-dimensional tensor with shape $[\,B,\ 2,\ R,\ \Theta,\ N_{k_y}\,]$, where $B$ denotes the mini–batch size. This tensor therefore represents the full three-dimensional, time-local turbulent state
$\phi(r,\theta,\alpha)$ resolved spectrally in $(r,\alpha)$ (i.e. respectively $k_x$ and $k_y$ wave numbers) and discretely in $\theta$.
In a typical CGYRO simulation, the radial spectral resolution is 
$R\sim\mathcal{O}(10^2$--$10^3)$, while the poloidal resolution is
$\Theta \gtrsim 24$, depending on the physics regime being modeled.  
As discussed later, we apply a physics-informed dimensionality reduction to the
radial Fourier modes based on a Lorentzian fit to the $k_x$ spectrum, thereby
retaining only the dynamically relevant portion of the radial support.
For the CGYRO dataset used in this work, the raw dimensions are
$(R, \Theta, N_{k_y}) = (324,\,24,\,16)$.

After applying the reduced-model spectral cropping, the input tensor supplied
to the encoder has the shape
\begin{equation}
    [B,\ 2,\ R_0,\ \Theta_0,\ N_{k_y}]
\end{equation}
where $R_0= a \,\Delta k_x, \quad 1<a<5$ of the
Lorentzian width in the radial spectrum.  
This reduced representation captures the essential turbulent features while
greatly improving computational efficiency.  
The dimensions introduced here will be used consistently throughout the
description of the U-Net--TCN architecture.

\subsubsection{Flux Targets}

CGYRO also provides species‐dependent turbulent fluxes:
\begin{equation}
    y(t_k) = 
    \big(Q_i(t_k),\, Q_e(t_k),\, \Gamma(t_k)\big),
\end{equation}
corresponding to the same temporal resolution as the stored $\Phi$ fields.
Training uses only the statistically steady saturated interval; early transient 
times are excluded. Calculation of additional transport channels such as impurity energy will be the subject of future work.

\subsubsection{Rationale for 3D Convolutions}

Three‐dimensional convolutions preserve the gyrokinetic structure of 
turbulence, which is not separable in $(r,\theta,\zeta)$:  
ballooning parity, eddy elongation, zonal‐flow shearing, and toroidal 
phase‐correlation all rely on full 3D resolution.  
The CNN learns spatial invariants including: radial correlation lengths, poloidal tilting and ballooning structure, toroidal coherence of eddies, local shear‐suppression patterns, streamer formation and break‐up. The Conv3D operator:
\bea
    (\mathrm{Conv3D} * \Phi)(k_x,k_y,\theta)
\eea
captures correlations in all three directions simultaneously.

\subsubsection{Training Windowing vs. Autoregressive Deployment}

An important methodological distinction arises between the temporal context used during training and the autoregressive forecasting regime in which the surrogate is ultimately deployed. During training, the temporal convolutional network is provided with a finite window of consecutive field snapshots $\{\Phi(t-T_c+1), \ldots, \Phi(t)\}$ and optimized to predict $\Phi(t+H)$ and the associated transport fluxes. From a strict information-theoretic perspective, this corresponds to a window-conditioned inference task rather than a pure step-ahead predictor, since the training loss is evaluated using ground-truth data across the full input window.

Crucially, however, the surrogate is evaluated and applied exclusively in autoregressive rollout mode, which enforces the operational constraints relevant for surrogate-based turbulence evolution. In this regime, the model predicts $\hat{\Phi}(t_0+H)$ using only ground-truth history up to $t_0$, after which its own predictions are recursively fed back as inputs to generate $\hat{\Phi}(t_0+2H)$, $\hat{\Phi}(t_0+3H)$, and so on. At no stage does the model access future ground-truth states during rollout; its long-horizon behavior is therefore determined entirely by the learned temporal dynamics and the stability of its internal representation under self-generated inputs.

From a generic machine-learning standpoint, one might expect a mismatch between window-based training and autoregressive deployment to lead to degraded rollout performance due to distribution shift, as the model transitions from clean ground-truth inputs to noisier self-predicted states. Nevertheless, our results (Figs.~6--10) demonstrate that the surrogate maintains excellent qualitative fidelity, preserves spectral structure, and reproduces transport statistics over rollouts extending beyond five turbulence autocorrelation times ($\sim1400$ steps). This suggests that the model has learned a robust representation of physically admissible turbulent states, enabling it to remain on a stable attractor under repeated autonomous evolution.

We hypothesize that for strongly physics-constrained systems such as gyrokinetic turbulence, window-based temporal training provides implicit regularization by encouraging temporal consistency and suppressing dataset-specific shortcuts. Long-horizon rollout stability thus reflects successful learning of a physics-constrained manifold supporting stable autoregressive dynamics.

\subsection{Autoregressive rollout evaluation}
\label{subsec:rollout}

To assess the dynamical fidelity of the surrogate model beyond one–step flux 
prediction, we perform an \emph{autoregressive rollout} test.  
Given an input window of length~$T_c$ (the temporal context), the model is 
initialized at time $t_0$ with the true CGYRO fields 
$\{\Phi(t_0 - (T_c-1)), \ldots, \Phi(t_0)\}$ and predicts the next field 
$\hat{\Phi}(t_0+1)$.  
This prediction is then recursively fed back into the model as input, replacing 
the true field, so that the model evolves the system forward for 
$N_{\mathrm{roll}}$ steps: $\Phi_{\mathrm{in}}(t+1) =\Phi_{\mathrm{pred}}(t+1)$. No teacher-forcing is applied during the rollout test itself; all predictions are generated in free–running mode.

\subsection{Training procedure}

The model is trained with Adam optimizer
\cite{kingma2014adam}, with learning-rate scheduling used to reduce the step size as the validation error saturates. Training is terminated using an early-stopping criterion to prevent overfitting and ensure stable convergence. The loss function is designed to combine multi-horizon, single-time flux prediction and field reconstruction losses together with their corresponding rollout consistency losses. details of the loss function calculation are presented in the appendix~\ref{app:loss}. 

Mini-batches are constructed from uniformly sampled temporal windows of length $T_c$ drawn from the available time series. Both input fields and target fluxes are normalized using statistics computed on the training subset only. All experiments are implemented in a modern deep-learning framework and utilize mixed-precision arithmetic to improve computational efficiency on accelerator hardware.

\subsection{Summary and novelty}

The proposed \GKFieldFlow model integrates:
\begin{enumerate}
    \item \textbf{A 3D spatial encoder} that extracts multiscale geometric features of the 
    turbulent potential across $(k_,k_y,\theta)$;
    \item \textbf{A dilated TCN} that learns the nonlinear temporal evolution of 
    gyrokinetic turbulence in latent space, enabling autoregressive rollouts; and
    \item \textbf{A dual-head architecture} where one head reconstructs $\Phi(t+H)$ at full 
    spatial resolution while the other predicts the associated nonlinear fluxes.
\end{enumerate}

This combination of volumetric convolutions, U-Net skip connectivity, and a TCN sequence 
model produces a surrogate that is both physically interpretable and capable of forecasting the 
multi-scale nonlinear evolution generated by CGYRO.

\begin{figure*}
     \centering
	   \begin{subfigure}[t]{0.6\linewidth}
		\includegraphics[width=1.0\linewidth]{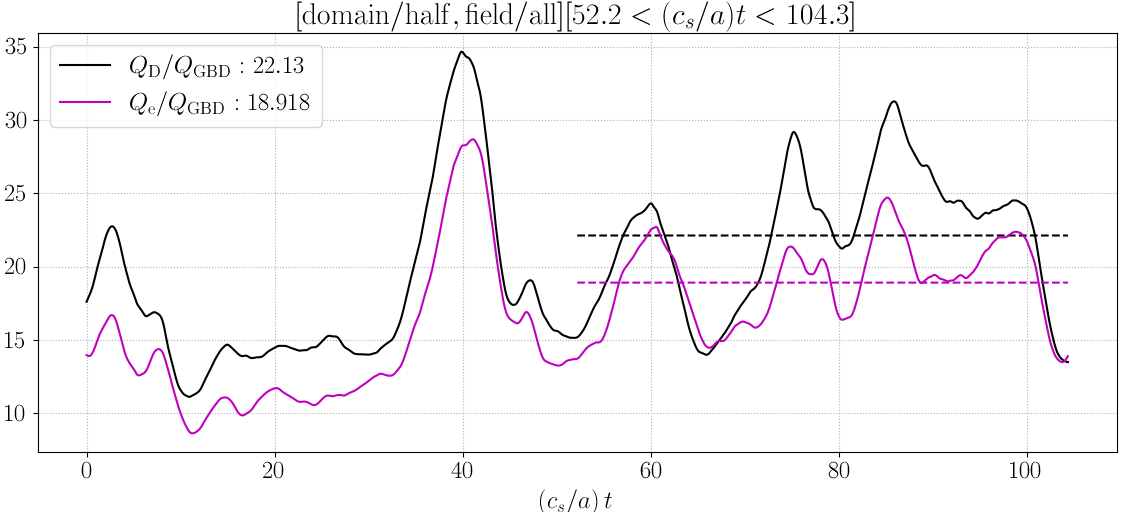}
        \caption{}
		\label{fig:flux_trace}
	   \end{subfigure}
	   \begin{subfigure}[t]{1.0\linewidth}
        \centering  
		\includegraphics[width=0.62\linewidth]{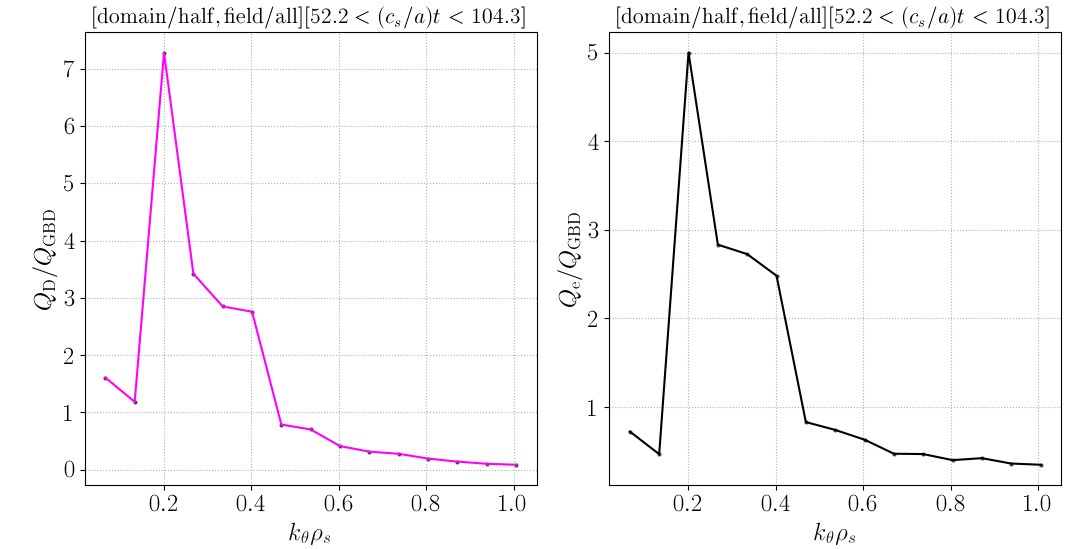}
        \caption{   }
		\label{fig:flux_dist}
	    \end{subfigure}
        \begin{subfigure}[t]{1.0\linewidth}
         \centering
		\includegraphics[width=0.6\linewidth]{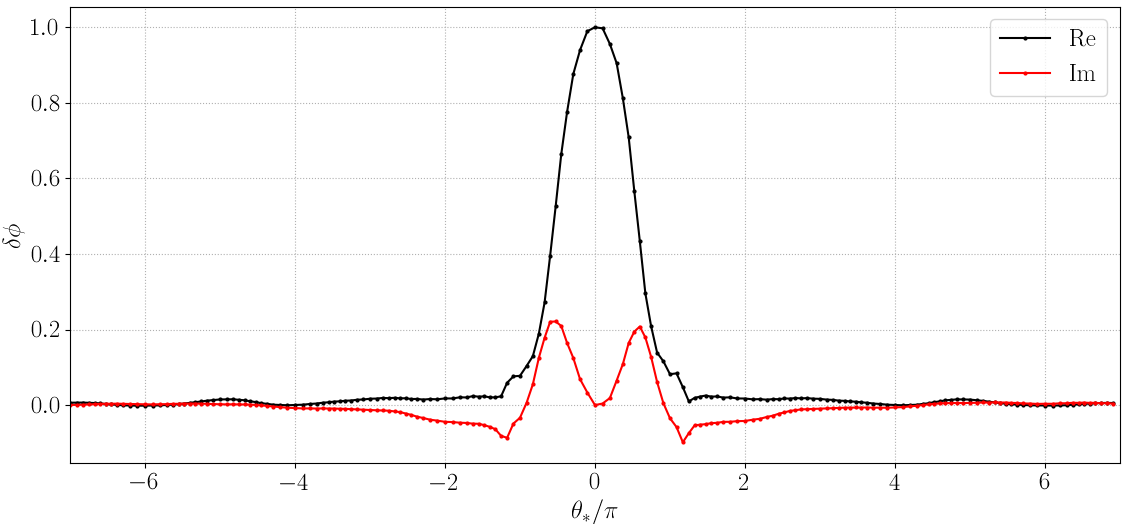}
        \caption{}
		\label{fig:lin_ball}
	    \end{subfigure}
	\caption{(a) Flux time traces of the test-case nonlinear CGYRO simulation, showing ion (deuterium), and electron energy, respectively $Q_D$ and $Q_e$ in gyroBohm normalized units calculated using nonlinear CGYRO  (b) Distribution of fluxes as a function of $k_y$ (c) linear eigenmode (from linear CGYRO) at the peak of the distribution $k_y=0.2$.}
	\label{fig:test_case}
\end{figure*}

\begin{figure*}[t]
\centering
\includegraphics[width=0.75\textwidth]{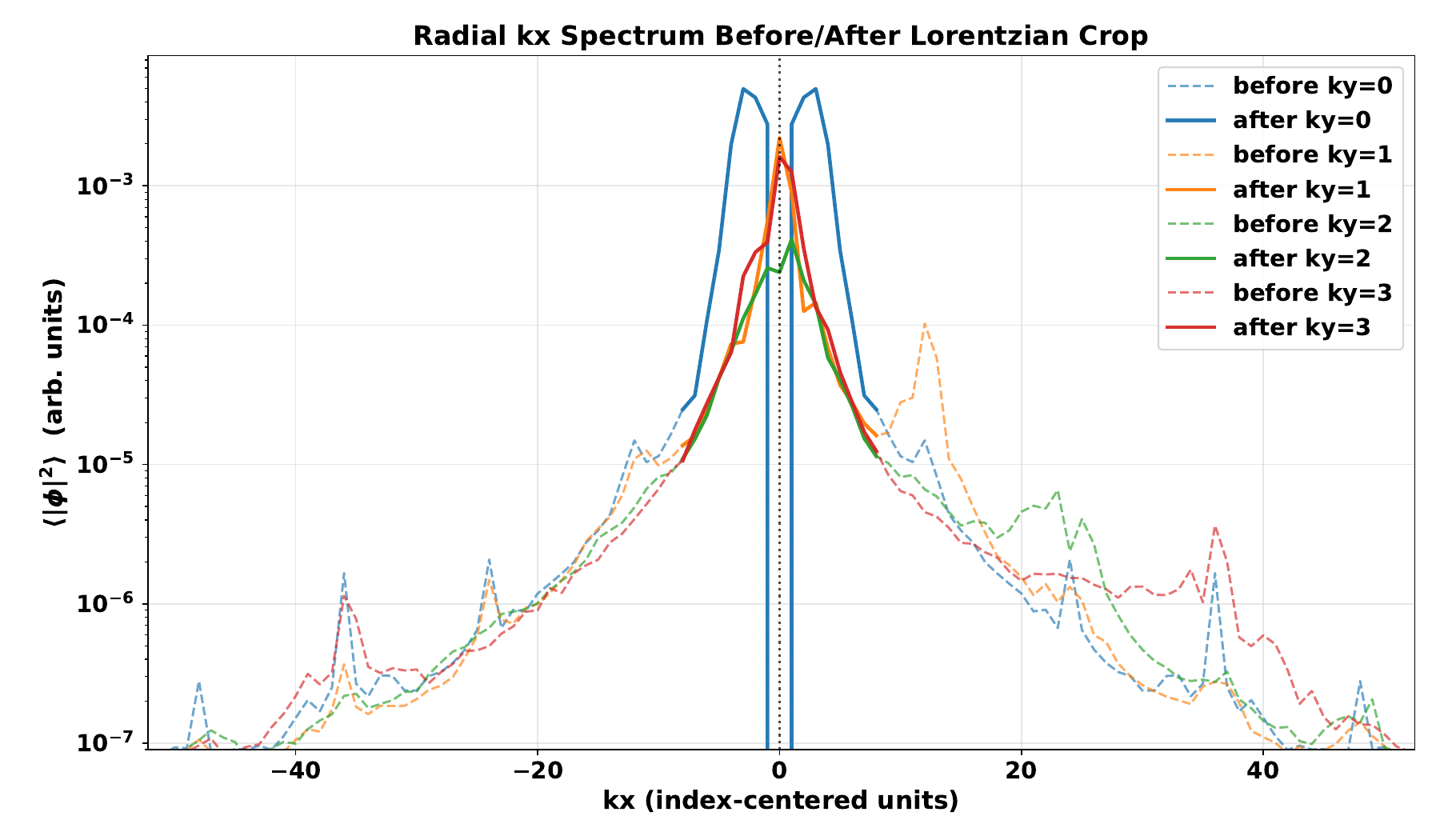}
\caption{Log plot of time-averaged radial distribution function of $\pm 50$ radial wave numbers around $k_x=0$ of the first 4 binormal modes($k_y=0$ is zonal). Solid curves show the cropped distribution for width of $2\Delta k_x$, where $\Delta k_x$ is calculated by the full-width-at-half-maximum (FWHM) criterion. Number of the kept cropped radial terms is $R_c=17$ from the original $R=324$ for this test example.}
\label{fig:Lorentzian}
\end{figure*}

\section{\label{crop}Dimensionality Reduction via Lorentzian Spectral Width Estimation}
\label{sec:lorentzian_crop}

A key objective of the present work is to construct a high–fidelity, 
computationally efficient surrogate model capable of predicting 
gyrokinetic turbulent fluxes using a significantly reduced set of input 
features. The raw CGYRO output contains several hundred radial Fourier 
modes in $k_x$ for each $(k_y,\theta)$, resulting in extremely large 
input tensors for any machine-learning model. This full dimensionality 
is needed in the CGYRO nonlinear simulations for accurate prediction of turbulent transport fluxes. However, 
gyrokinetic turbulence exhibits a well-known spectral structure: only 
a compact region of the radial wavenumber spectrum carries the 
dynamically relevant fluctuation energy. This observation motivates the use of a physics-based dimensionality-reduction procedure in which the 
dominant $k_x$ region is identified and retained, while high-$|k_x|$ 
modes---which contribute negligibly to transport but significantly 
inflate data size---are systematically discarded.

\subsection{Lorentzian Structure of Drift-Wave Turbulence Spectra}

The radial spectrum of electrostatic potential fluctuations,
\bea
   S(k_x) \equiv \left\langle |\phi(k_x,k_y,\theta_{\mathrm{out}})|^2 
   \right\rangle_t ,
\eea
is sharply peaked around $k_x=0$ and exhibits tails that approximate a 
Lorentzian or modified-Lorentzian envelope. This spectral form is a 
direct consequence of the underlying real-space turbulence structure. 
Drift-wave turbulence is characterized by exponentially decaying radial 
two-point correlation functions, driven by nonlinear eddy mixing and 
the decorrelating effect of sheared flows. Exponential correlations 
imply Lorentzian power spectra in Fourier space, a classical result of 
statistical theory \cite{LandauLifshitzStatPhys}. 

Moreover, zonal flows act to shear turbulent eddies, causing spectral 
broadening that further reinforces a Lorentzian-like shape 
\cite{Diamond2005}. Theoretical treatments of drift-wave turbulence 
show that exponential radial decorrelation, when combined with streamer 
dynamics, naturally leads to Lorentzian spectra in $k_x$ 
\cite{horton1999}. Modern nonlinear gyrokinetic simulations, including 
studies of ITG, TEM, ETG, and microtearing turbulence, confirm that 
the saturated-state radial spectra are well fit by Lorentzian envelopes 
across a wide parameter regime \cite{Hatch2011}. This robust physical 
basis provides a natural framework for using Lorentzian widths as 
quantitative measures of the turbulence correlation scale.

\subsection{Test Case: Nonlinear CGYRO Simulation}
\label{sec:testcase}

Results from the representative nonlinear gyrokinetic simulation used in this
study are shown in Fig.~\ref{fig:test_case}. Panels~(a)--(c) display the
turbulent heat flux time traces, the corresponding binormal spectral
distribution, and the dominant linear eigenmode structure. The case is an
ion-scale simulation with the base wave number $\Delta k_y\rho_s =0.067 $ and $N_{k_y}=16$ binormal modes, which results in the total $k_y$ spectral range of $\approx 1.0$. The simulation is performed for the 
normalized minor radius \(r/a = 0.8\) inside an L-mode discharge in the DIII-D
tokamak. The numerical resolution consists of \(N_r = 324\) radial Fourier terms and 
\(N_\theta = 24\) poloidal grid points. 
Explicit transient removal has been applied, and the time window shown
corresponds to a saturated turbulent state.

The dominant heat flux \(Q_D\) is carried in the ion (deuterium) channel.
Linear CGYRO analysis indicates that the most unstable mode appears near the
peak of the \(k_y\)-spectrum in Fig.~\ref{fig:test_case}(b), propagates in the
ion-diamagnetic direction, and exhibits a ballooning-parity structure with a
pronounced outboard-midplane peak [Fig.~\ref{fig:test_case}(c)]. These features
are consistent with toroidal ion-temperature-gradient (ITG) instability,
suggesting that ITG turbulence is the primary drive mechanism in this regime.
As ITG turbulence is predominantly electrostatic, the present study focuses on
predicting the electrostatic potential \(\Phi\) as the input field for the
surrogate. Extension to electromagnetic simulations involving
\((\Phi, A_\parallel)\) will be explored in future work.

This simulation serves as the reference dataset for training and evaluating
\GKFieldFlow{} throughout the remainder of this article.

\subsection{Radial Spectral Cropping in $k_x$}

To reduce the dimensionality of the input fields while retaining the dominant
dynamically relevant content, we perform a Lorentzian-like
width estimate based on the full-width-at-half-maximum (FWHM) criterion of the global radial spectrum $S_{\mathrm{global}}(k_x)$. If the half-maximum method is inconclusive (e.g.\ plateau or
shoulder structure), a secondary estimate based on the second spectral moment
around the peak is used as a fallback. For each dataset, we compute a global radial spectrum
\begin{equation}
S_{\mathrm{global}}(k_x)
= \big\langle \,|\Phi|^2 \,\big\rangle_{t,\theta,k_y},
\end{equation}
defined as the time- and angle-averaged energy density of the potential
fluctuations. This provides a device- and case-independent measure of the radial
mode distribution commonly used to characterize turbulent structure in
gyrokinetic simulations~\cite{told2015,howard2016,CGYRO}. We identify the principal peak of $S_{\mathrm{global}}(k_x)$ and estimate an
effective spectral width $\Delta k_x$ using the FWHM criterion. A symmetric crop radius $R_c$ is then
chosen as a controlled multiple of this width,
\begin{equation}
k_x \in [-R_c,\,R_c], \qquad R_c = \alpha\,\Delta k_x , 
\quad \alpha = \mathcal{O}(1\!-\!4),
\end{equation}
where the scaling factor $\alpha$ determines how much of the tail structure in
$S_{\mathrm{global}}(k_x)$ is retained.

This produces a truncated field $\Phi_{\mathrm{crop}}$ in which the majority of
the physically relevant turbulent energy is preserved while removing
high-$k_x$ content that contributes minimally to flux prediction but increases
computational cost. The same crop is applied uniformly for all $k_y$ to maintain
tensor consistency and avoid biasing individual spectral channels. In this way,
the dominant radial scales are retained without case-by-case tuning, providing a
reliable reduction in input dimensionality.

An example for such procedure is shown in Fig.~\ref{fig:Lorentzian} which shows the log-plot of the timeand $\theta$ averaged radial distribution for the intensity of the first 4 binormal modes($k_y=0$ is the zonal flow which does not contain $k_x=0$). Dashed curves show the actual distribution which spans to $\pm 192$; here we zoom in and only show $\pm 50$ terms. The solid curves show the cropped distribution for the width of $2 \Delta k_x$ where $\Delta k_x$ is calculated using the FWHM method. With this cropping method, the radial terms kept is reduced from $R_c=324$ to only $R_c=17$. This choice captures nearly all fluctuation energy in the physically relevant 
portion of the spectrum while excluding high-$|k_x|$ tails that contain 
at most a few tenths of a percent of the total energy. These tails are 
dominated by small-amplitude structures that do not significantly 
contribute to turbulent transport but substantially increase the number 
of input modes.

\subsection{Impact on Surrogate Model Efficiency}

Reducing the number of radial modes directly lowers the dimensionality 
of the model input tensor and thereby decreases the memory footprint, 
training time, and inference latency of the \GKFieldFlow surrogate 
model. High-dimensional inputs can lead to overfitting and degrade 
model stability, especially when training data are limited. By 
systematically removing only those modes that are physically irrelevant, 
the Lorentzian-based cropping procedure improves model generalization 
and efficiency without sacrificing fidelity to the underlying 
gyrokinetic dynamics. The resulting dataset is both more compact and 
more physically interpretable, making it particularly well suited for 
fast, predictive surrogate modeling of turbulent transport.

\section{\label{depth}Effect of U\textendash Net Depth and Radial Spectral Cropping on Predictive Accuracy}
\label{sec:layers_vs_rc}

To quantify how architectural depth and input spectral resolution influence the 
performance of the spatial encoder, we performed a two\textendash dimensional 
scan over (i) the maximum number of U\textendash Net downsampling levels and 
(ii) the number of retained radial spectral modes~$R_c$.  
All models in this study employed a consistent channel-scaling strategy across
spatial resolution levels, with feature dimensionality increasing at deeper
encoder stages. For each combination of architectural depth and spatial context
parameters, a complete \GKFieldFlow{} model was trained to convergence and
evaluated using the normalized root-mean-square error (NRMSE), averaged over the
three turbulent transport channels $(Q_i, Q_e, \Gamma)$.
The rollouts were initialized from an ensemble of uniformly distributed start 
times and propagated {560 time steps, which, using a stride of~2 in~$t$--responds
to approximately one turbulence autocorrelation time ~$T_{\mathrm{ac}}\approx 9 \: a/c_s$ for the tested CGYRO simulation ($a$ is the tokamak plasma minor radius and $c_s= \sqrt{T_e/m_D}$ is the ion sound speed, $T_e$ is electron temperature, and $m_D$ is the deuterium mass). Furthermore, we used a stride of 4 to reduce the resulotion in the poloidal direction. Therefore, even before the radial cropping the temporal and poloidal reduction with strides has reduced the size of input data by a factor of 8. 
\begin{table*}[t]
\centering
\caption{Best--epoch single--time inference errors for the scan over 
U\textendash Net depth (\texttt{levels}) and the number of retained radial Fourier 
modes $R_c$.  Bold entries indicate improvements with increase in depth of $\ge20\%$.
for each metric.}
\label{tab:single_time_errors}
\renewcommand{\arraystretch}{1.20}
\begin{tabular}{c|c|c c|c c c c}
\hline
levels & $R_c$ & Train Loss & Val Loss & RMSE$_{Q_i}$ & RMSE$_{Q_e}$ & RMSE$_{\Gamma}$ & RMSE$_{\Phi}$ \\
\hline
1 & 11 & 0.00923 & 0.00169 & 0.03096 & 0.02898 & 0.03075 & 0.02493 \\
2 & 11 & \textbf{0.00473} & \textbf{0.00084} & \textbf{0.01882} & \textbf{0.01669} & 0.02646 & \textbf{0.01770} \\
\hline
1 & 11 & 0.00923 & 0.00169 & 0.03096 & 0.02898 & 0.03075 & 0.02493 \\
2 & 11 & \textbf{0.00473} & \textbf{0.00084} & \textbf{0.01882} & \textbf{0.01669} & 0.02646 & \textbf{0.01770} \\
\hline
1 & 17 & 0.01449 & 0.00163 & 0.02503 & 0.02024 & 0.02862 & 0.02839 \\
2 & 17 & \textbf{0.00789} & \textbf{0.00111} & \textbf{0.01940} & 0.02322 & 0.02442 & \textbf{0.02193} \\
3 & 17 & \textbf{0.00589} & 0.00118 & 0.02337 & 0.02903 & 0.02733 & 0.01927 \\
\hline
1 & 33 & 0.01440 & 0.00181 & 0.02196 & 0.02552 & 0.03374 & 0.02904 \\
2 & 33 & \textbf{0.00880} & \textbf{0.00133} & 0.02183 & 0.02564 & \textbf{0.02238} & 0.02513 \\
3 & 33 & 0.00730 & 0.00112 & 0.02290 & 0.02557 & 0.02241 & \textbf{0.02109} \\
\hline
1 & 49 & 0.01370 & 0.00130 & 0.02107 & 0.02064 & 0.02741 & 0.02464 \\
2 & 49 & 0.01110 & 0.00142 & 0.02086 & 0.02662 & 0.02388 & 0.02607 \\
3 & 49 & 0.01587 & 0.00342 & 0.03785 & 0.04736 & 0.04493 & 0.03494 \\
\hline
\end{tabular}
\end{table*}

\subsection{Single-time inference performance}
A first measure of the fidelity of the surrogate model is its ability to infer
the heat and particle fluxes, as well as the turbulent electrostatic
potential, at a single target time given the input window of $T_c$
consecutive gyrokinetic fields.
In machine-learning applications to turbulence, RMSE values of order
$10^{-1}$ (normalized units) or a few percent relative error in physical
units are generally considered strong performance for high-dimensional
multiscale dynamical systems \cite{mohan2018deeplearningbasedapproach,mishra2023estimatesgeneralizationerrorphysics}.
By this standard, our model demonstrates excellent accuracy: normalized
flux RMSE values fall in the range $1.1\times10^{-2}$–$2.0\times10^{-2}$,
with corresponding physical errors of the order $5\times10^{-2}$–$1.2\times10^{-1}$.
The electrostatic potential achieves normalized errors at the level of
$3\%$–$4\%$.

\paragraph{Single–time inference performance across U\textendash Net depth and radial resolution}

Table~\ref{tab:single_time_errors} reports the best--epoch errors for
single--time flux inference across the scanned combinations of U\textendash Net
depth and radial spectral truncation~$R_c$. The values of $R_c=11,17,33$ and $49$
correspond to $1.25\times$, 2$\times$, 3$\times$, and 4$\times$ the Lorentzian
width $\Delta k_x$.
From the perspective of standard machine--learning benchmarks, the obtained
NRMSE values are remarkably low: for many configurations the flux errors fall in
the range $1.8\%\text{--}3.0\%$, well within what is typically considered
``good'' to ``very good'' performance for nonlinear regression on chaotic
dynamical systems.
The potential--field errors, often below $2.5\%$, further indicate that the
reduced spatial representation retains the dominant physical content required
for accurate flux inference.

Several clear trends emerge.
First, for strongly cropped inputs ($R_c = 11$ and $17$), increasing the
U\textendash Net depth from one to two levels yields a substantial reduction in
flux RMSE across all species.
This demonstrates that deeper hierarchical encoding is beneficial when only the
largest--scale radial features are present.
A third level was not applied for $R_c=11$ due to collapse of the $\Theta$
dimension, but for $R_c = 17$ it provides only marginal additional improvement.

For intermediate resolution ($R_c = 33$), performance improves only for the last
two columns when increasing from one to two levels and continues to improve
slightly at three levels, although the gains are modest.
This regime appears to lie near a transition in which the encoder's
downsampling becomes increasingly costly but not yet detrimental.

In contrast, for high resolution ($R_c = 49$) the three--level model performs
significantly worse, with flux errors nearly doubling relative to the two--level
network.
Here, the spatial downsampling becomes excessively compressive: fine radial
structure and short correlation lengths cannot be preserved through the deeper
bottleneck, leading to degraded inference accuracy.

It is important to note that differences at the $10^{-3}$ level in RMSE lie
within the expected stochastic variability of neural--network training.
Thus, only the \emph{large} performance differences---such as the strong
improvement from one to two levels at low~$R_c$ or the sharp degradation of the
three--level model at $R_c=49$---should be interpreted as statistically
meaningful.

Overall, these results reinforce a central conclusion of this work: reduced
representations with very few retained radial modes ($R_c=11$--17) perform
surprisingly well.
This validates the physics--motivated reduced modeling strategy adopted here.
Refinements to enable deeper encoders at high spectral resolution constitute an
important direction for future research.

% -------------------------------------------------------------------
\subsection{Autoregressive rollout performance}

Single--time inference metrics provide a baseline measure of how well the 
network approximates the nonlinear mapping 
$\Phi(t) \mapsto (Q_i, Q_e, \Gamma)(t+\Delta t)$.
Such one--step losses are widely used in the literature as indicators of 
pure function--approximation capability.  
However, for autoregressive surrogates of turbulent dynamics, one--step accuracy 
is not sufficient to predict rollout performance.  
Small phase or amplitude errors in $\Phi$ may amplify under chaotic 
time evolution, even when the one--step flux RMSE is small.  
Accordingly, models that appear similar in single--step accuracy may exhibit 
large differences during rollouts.  
Our results clearly demonstrate this: architectures such as $(2,17)$ and
$(3,17)$ show comparable single--time flux RMSE, yet $(3,17)$ yields
significantly improved long--horizon behavior due to better reconstruction of
the multi--scale structure of $\phi$.  

Thus, assessing surrogate fidelity requires both single--time tests 
and rollout diagnostics, the latter being more sensitive to the stability of
the learned latent dynamics.

To evaluate long-term dynamical fidelity, we perform autoregressive
rollouts: the model predicts $\Phi(t+H)$ and the associated fluxes,
then feeds its own inference back as input.  This process continues for
$N_{\mathrm{roll}}$ steps without access to the true CGYRO data.

Figure~\ref{fig:LayersvR} displays the resulting normalized RMSE (NRMSE) for the rollouts of (a) turbulent fluxes, (b) turbulence potential $\Phi,(k_y>0)$ and (c) zonal flow potential $\Phi,(k_y=0)$, as a function of 
    the number of retained radial spectral modes $R_c$ and the maximum 
    U\textendash Net depth. The scan dimensions are the same as for table~\ref{tab:single_time_errors}, i.e. levels:(1,2,3) vs $R_c:(11,17,33,49)$. This error landscape exhibits a nontrivial relation between the depth of
the spatial U\textendash Net encoder and the number of retained radial spectral modes~$R_c$.    
In the strongly cropped regime ($R_c=11$ and $17$), the turbulent potential contains only 
large\textendash scale radial structure.  
Here, increasing the number of encoder levels yields a monotonic improvement in prediction
accuracy: deeper U\textendash Nets enlarge the spatial receptive field and
provide a more expressive hierarchical representation of $\phi(\textbf{r},t)$.
For $R_c=11$, only two downsampling stages are performed, since a third stride\textendash 2
reduction would collapse the $\Theta$ dimension; nevertheless, the transition from one to
two levels significantly reduces the NRMSE.

The behavior changes qualitatively for larger spectral bandwidths ($R_c=33$ and $49$).
In this high\textendash resolution regime, the three\textendash level U\textendash Net performs 
\emph{worse} than the two\textendash level model.  
Retaining more radial modes introduces finer spatial scales and shorter radial
correlation lengths; a deep sequence of stride\textendash 2 downsamplings 
over\textendash compresses these features, blurring or erasing information that is crucial for 
accurate flux prediction.  
The poorer performance of the three\textendash level network at large $R_c$ indicates that
the encoder architecture must adapt as the input resolution increases.

These observations establish a resolution\textendash dependent design principle:
\emph{coarsely cropped inputs benefit from deeper hierarchical encoders, whereas
high\textendash resolution fields require shallower U\textendash Nets unless architectural 
modifications are introduced to prevent over\textendash compression}.  
It is important to emphasize that the objective of the present work is not to
capture the full spatial resolution of the gyrokinetic potential, but rather to
develop a \emph{reduced} surrogate model that performs as well as possible with
the smallest amount of information. Objectively comparing the all the settings tested as shown in Fig.~\ref{fig:LayersvR} the best performing configurations are first, $R_c=17$ radial terms with 3 U-Net levels, followed by $R_c=11$ radial terms with 2 U-Net levels.       
In this context, the results of Fig.~\ref{fig:LayersvR} are significant:
a model retaining only $R_c = 11$ radial Fourier modes performs comparably to,
or in some cases better than, models using $R_c = 17$ or even $R_c = 33$,
irrespective of whether two or three U\textendash Net levels are employed.
This demonstrates that much of the information essential for accurate flux
prediction is contained in the lowest\textendash order radial spectrum, and thus
validates the reduced\textendash resolution modeling strategy adopted here.

Interestingly, the single--time inference metrics (training loss, validation 
loss, and one--step RMSE of $Q_i$, $Q_e$, and $\Gamma$) do not correlate 
directly with the long--horizon rollout error.  
Models such as $(2,17)$ and $(3,17)$ both appear among the top performers 
in one--step flux inference, yet their rollout behavior differs 
substantially.  

As illustrated in Figs.~\ref{fig:LayersvR}(a)--(c), the decisive factor for 
rollout accuracy is not the single--time flux error but the model’s 
ability to reconstruct the turbulent potential $\phi$ across both the 
$k_y=0$ (zonal--flow) and $k_y>0$ (drift--wave) channels.  
The $(3,17)$ model, despite having slightly worse one--step flux RMSE than 
$(2,17)$, yields consistently lower NRMSE in both $\Phi(k_y=0)$ and 
$\Phi(k_y>0)$, indicating improved phase coherence, reduced error 
amplification, and greater stability of the learned temporal dynamics.  

This highlights a key principle of autoregressive turbulence modeling: 
rollout performance is governed primarily by the stability and accuracy of 
the latent dynamical representation, rather than by single--time flux 
inference accuracy.

While deeper U\textendash Nets and alternative encoder designs may enable
high- resolution inputs ($R_c \gtrsim 33$) to be exploited more
effectively, such refinements fall outside the scope of the present study.
Instead, we focus on establishing the effectiveness of a compact, 
physics\textendash motivated reduced representation.  
The development of enhanced architectures tailored to higher spectral bandwidth will be the subject of future work.
Together, these results underscore that optimal encoder depth must be matched to the
physical content of the input fields, and that deeper networks can be made effective at 
high spectral resolution only when supported by appropriate architectural modifications.

\begin{figure*}

	   \begin{subfigure}[t]{0.45\linewidth}

		\includegraphics[width=1.0\linewidth,trim={5.5cm 1.0cm 3cm 1cm},clip]{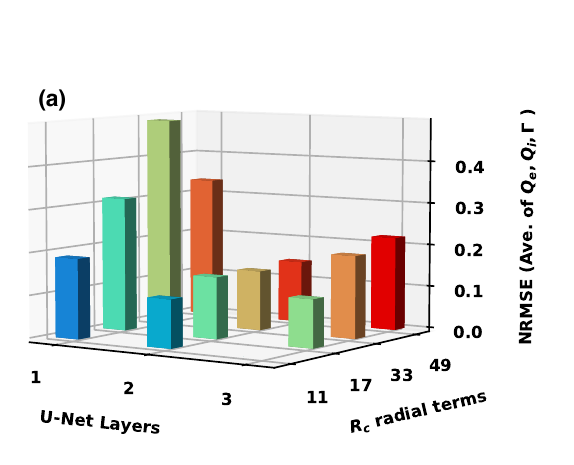}
        \caption{Turbulent fluxes }
		\label{fig:subfigL1}
	   \end{subfigure}
	   \begin{subfigure}[t]{0.45\linewidth}

		\includegraphics[width=1.0\linewidth,trim={5.5cm 1.0cm 3cm 1cm},clip]{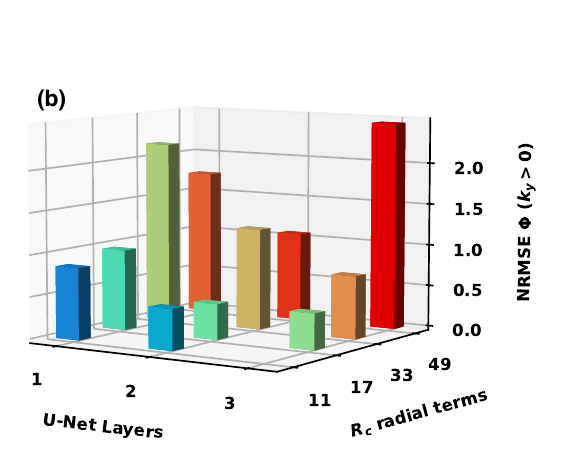}
        \caption{Turbulence potential $\Phi,(k_y>0)$   }
		\label{fig:subfigL2}
	    \end{subfigure}
        \begin{subfigure}[t]{0.45\linewidth}

		\includegraphics[width=1.0\linewidth,trim={5.5cm 1.0cm 3cm 1cm},clip]{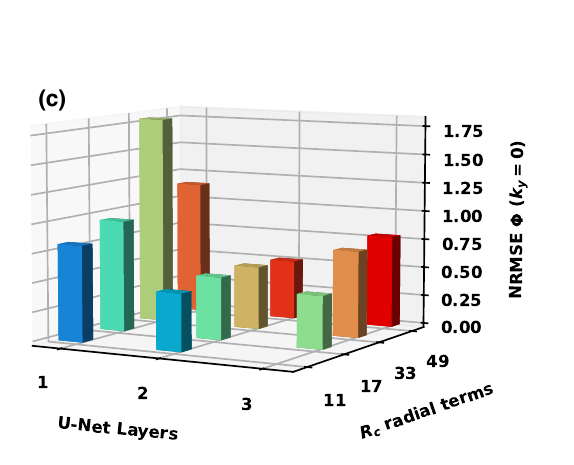}
        \caption{Zonal flow potential $\Phi,(k_y=0)$ }
		\label{fig:subfigL3}
	    \end{subfigure}
	\caption{ Normalized RMSE (NRMSE) for the rollouts of (a) turbulent fluxes (b) turbulence potential $\Phi,(k_y>0)$, and (c) zonal flow potential $\Phi,(k_y=0)$, as a function of 
    the number of retained radial spectral modes $R_c$ and the maximum 
    U\textendash Net depth.  
    For $R_c=11$, only two downsampling levels are possible due to collapse 
    of the $\Theta$ dimension.  }
	\label{fig:LayersvR}
\end{figure*}

\section{\label{scan_tc}Optimal Temporal Context $T_c$ for TCN}

A central question in designing autoregressive gyrokinetic surrogates is:
\emph{How much temporal history does the model truly need in order to 
accurately advance the turbulent state?} To determine the effective temporal memory required by the neural surrogate to 
accurately reconstruct both the turbulent dynamics and the associated heat and 
particle fluxes, we carried out a systematic scan over the temporal input window 
$T_c$.  
For each value of $T_c$ we trained a separate model using the same U-Net--TCN 
architecture, chose the best epoch out of 5 different runs for every single $T_c$, and evaluated the ensemble-averaged rollout error over 
60 rollout initial times, with a fixed rollout length of 280 steps which for a stride 4 time sampling used here corresponds to one turbulence aoutocorrelation time for this test case ($280\times4\times dt \approx 9 a/c_s$, where $dt=0.008$ in the CGYRO test-case). \par

\begin{figure}[t]
\centering
\begin{subfigure}{0.92\linewidth}
\centering
\includegraphics[width=\linewidth]{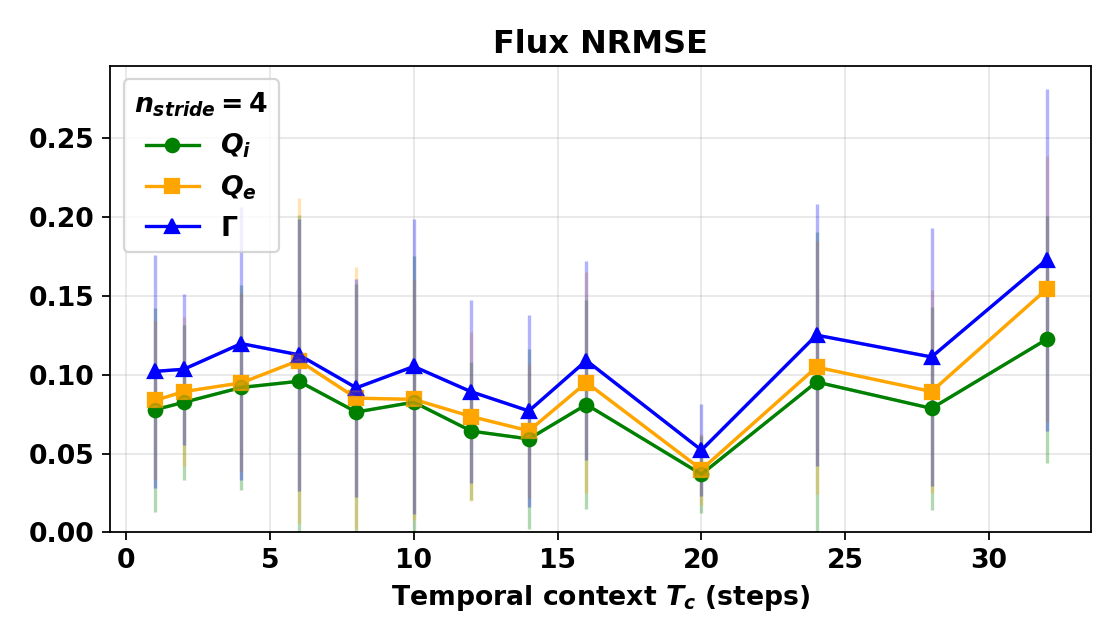}
\caption{\label{fig:Tc_scan_a}}
\end{subfigure}
\hfill
\begin{subfigure}{0.9\linewidth}
\centering
\includegraphics[width=\linewidth]{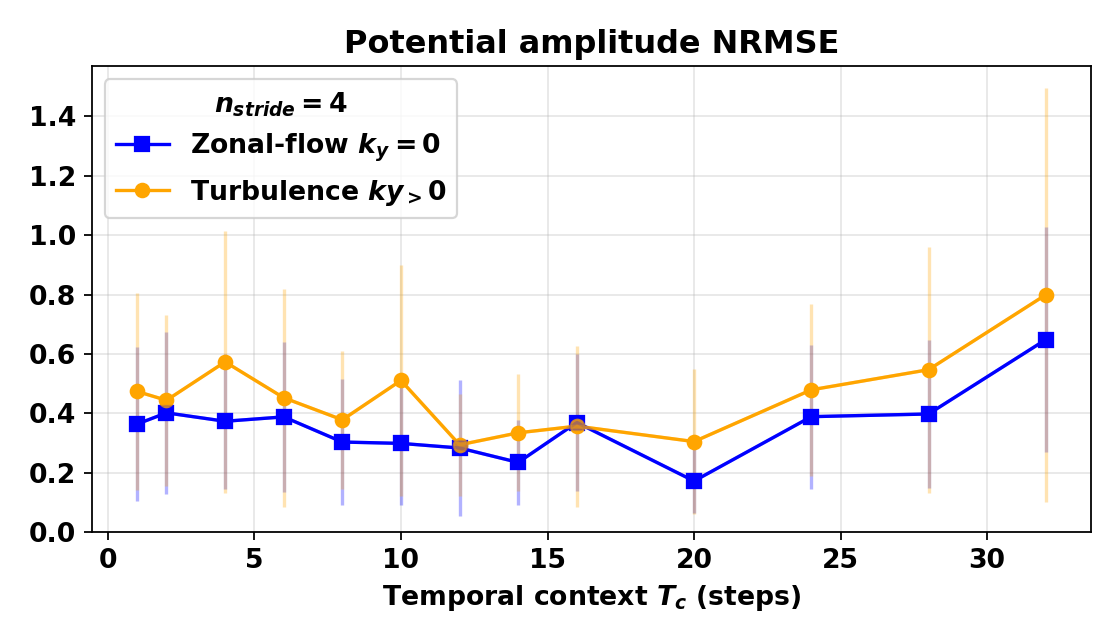}
\caption{\label{fig:Tc_scan_b}}
\end{subfigure}
\begin{subfigure}{1.0\linewidth}
\centering
\includegraphics[width=\linewidth]{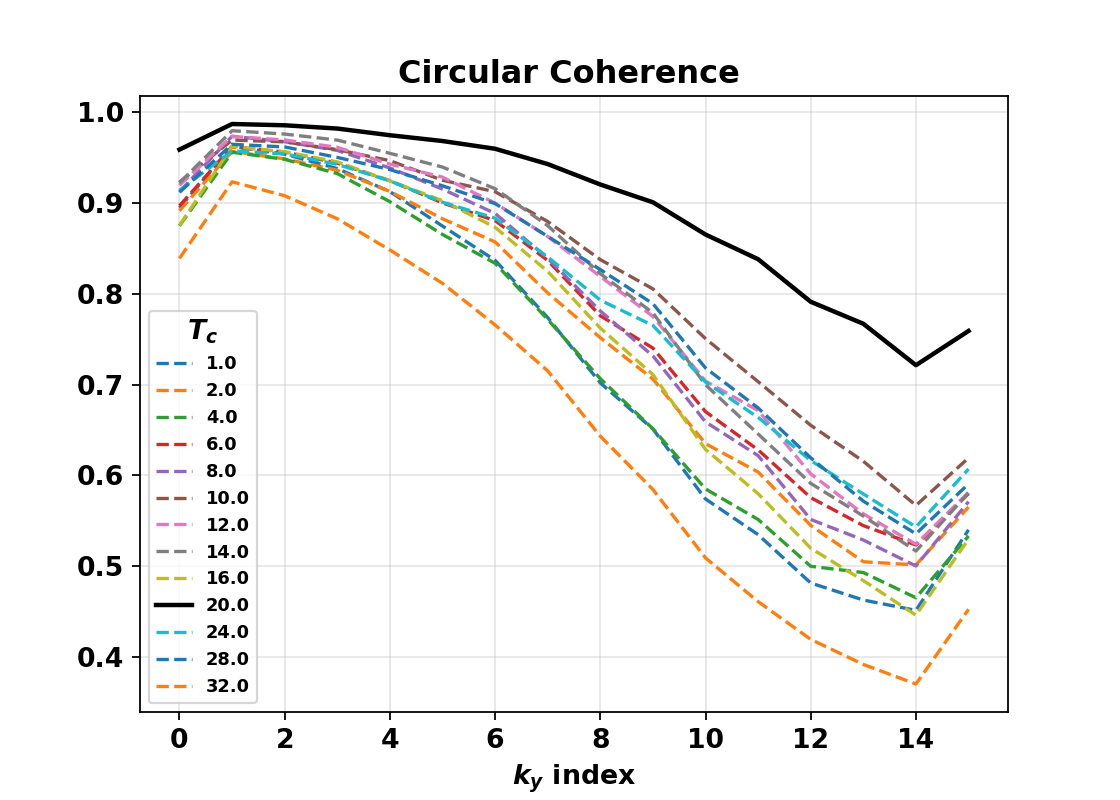}
\caption{\label{fig:Tc_scan_c}}
\end{subfigure}
\caption{\label{fig:Tc_scan}(a) Flux NRMSE  for $Q_i$, $Q_e$, and $\Gamma$, and (b) potential amplitude NRMSE for zonal-flow and turbulent components versus temporal context $T_c$. Error bars represent $\pm 1$ standard deviation across $N=40-60$ rollout initializations. (c) Circular coherence versus $k_y$ spectrum for the tested tempral contexts, showing $T^*_c=20$ has the highest coherence. As seen in (a) and (b), $T^*_c$ produces the minimum NRMSE for fluxes and turbulence amplitudes.}

\end{figure}

Using this framework, we examine the dependence of prediction accuracy  on the temporal context length $T_c$. The results are shown in Fig.~\ref{fig:Tc_scan}, which summarizes the effect of varying the temporal context
window $T_c$ in the \GKFieldFlow{} model evaluated
over rollouts spanning approximately one turbulence autocorrelation time
$T_{\rm ac}$. Three diagnostics are shown: (i) the
normalized RMSE of the predicted turbulent fluxes $(Q_i,Q_e,\Gamma)$, (ii) the
NRMSE of the potential amplitude for zonal-flow $(k_y=0)$ and turbulence
$(k_y>0)$ components, and (iii) a circular-coherence metric that measures the
phase alignment between predicted and true Fourier modes. Circular-coherence values  $C_\phi \approx 1$ indicate accurate reproduction of the turbulent phase structure by the model (see appendix~\ref{app:coh} for details on the derivation of $C_\phi$). Together, these
quantities characterize both the predictive accuracy and the structural
consistency of the autoregressive surrogate.

The $T_c=1$ case in Fig~\ref{fig:Tc_scan} corresponds to a purely instantaneous spatial surrogate, in which predictions are conditioned only on a single field snapshot with no temporal context. This setting involves no access to past or future states and therefore serves as a clean baseline for quantifying the added value of temporal windowing. Although no temporal history is provided for this case, a single turbulent snapshot encodes strong cues about transport through its spatial structure and amplitude. Since the dynamics are constrained to a narrow set of physically admissible states, the surrogate effectively learns to correct or stabilize each snapshot toward this set, rather than explicitly modeling time evolution. Hence, the $T_c=1$ case yields competitive performance despite the absence of temporal memory.

Across all diagnostics, a consistent trend emerges. As $T_c$ increases,
the normalized error decreases within the bounds of uncertainty, indicating that the TCN benefits from additional context and is able to extract more predictive structure from the time history of the latent states. This trend continues until an optimal
window is reached at approximately
\begin{equation}
T_c^\star \simeq 20 \;\;\text{(in model time steps)}, 
\end{equation}
beyond which the error begins to rise again. At this minimum, the flux NRMSE is
reduced by nearly $\sim 50\%$ relative to the $T_c=1$ configuration, and both
zonal-flow and non-zero $k_y$ components of the potential achieve their best
reconstruction fidelity. Importantly, Fig.~\ref{fig:Tc_scan_c}  shows that the circular coherence remains near unity
for low and intermediate $k_y$ in this regime and has higher coherence of all the other $T_c$ values. This indicates that the model
maintains phase alignment and avoids rapid decorrelation during rollouts.

The emergence of a minimum at $T_c^\star$ reflects the balance between two
competing effects. Larger windows provide more temporal information and
effectively lengthen the model's usable memory, improving predictions on
timescales shorter than the turbulence decorrelation time. However, as $T_c$
increases further, the added history predominantly consists of decorrelated turbulent fluctuations whose phases are no longer predictive of transport, increasing the effective learning complexity and lead to diminished generalization performance. For sufficiently
large $T_c$, this added variability acts as noise during training and
increases the difficulty of optimization, which manifests as a degradation in
performance for $T_c \gtrsim 20$. The resulting non-monotonic trend is therefore
interpreted as a competition between \emph{information gain} from extended
context and \emph{error accumulation} from noisy or weakly correlated inputs.

\begin{comment}
It is notable that the optimal memory window $T_c^\star$ is significantly
smaller than the turbulence autocorrelation time, indicating that the surrogate
learns a reduced effective memory: the network does not need to observe a full
autocorrelation epoch to make accurate predictions. Instead, it extracts the
shorter causal dependencies that are most relevant for the local temporal
evolution of the fields and the corresponding fluxes. This observation is
consistent with the improved flux prediction at $T_c^\star$ and the sustained coherence across moderate forecast horizons.

Because the network does not observe the full gyrokinetic state (e.g.\ velocity-space structure, kinetic moments, or electromagnetic fields), the evolution is not Markovian from the model's perspective. Instead, the surrogate sees only a low-dimensional projection of the true system, and accurate next-step prediction requires a finite history of past inputs. The optimal temporal context $T_c$ therefore reflects an \emph{effective learned memory time}: the window over which hidden phase-space information leaves correlated signatures in the observable fields. In practice, the network uses this short window to reconstruct the missing dynamics well enough to predict future fields and fluxes.
\end{comment}

Overall, these results indicate that the TCN is effectively exploiting short-range temporal dependencies in the gyrokinetic fields, while excessive temporal context introduces nuisance information that overwhelms the predictive signal. The resulting U-shaped dependence of error on $T_c$ is therefore consistent with a finite informational memory and a bias--variance tradeoff inherent to learning chaotic turbulent dynamics from snapshot-based data.

\section{\label{longh}Long-Horizon Rollout Stability at Optimal Temporal Context}

\subsection{Motivation for rollout testing}

While short-horizon inference accuracy is a necessary benchmark for any surrogate model, it is not sufficient to establish whether the model has learned a physically meaningful dynamical representation. In chaotic systems such as gyrokinetic turbulence, errors can remain small over short windows while still leading to rapid divergence under free-running evolution. It is therefore essential to assess the stability, coherence, and predictive fidelity of the surrogate under long-horizon rollouts, where the model evolves autonomously without access to ground-truth future states.

In this section, we evaluate the long-time rollout behavior of the \GKFieldFlow surrogate trained with temporal context $(T^*_c = 20)$, which was identified in Sec.~\ref{scan_tc} as the optimal context length for stride-4 temporal sparsification. Notably, this context length is much shorter than the turbulence autocorrelation time $(T^*_c \ll T_{\mathrm{ac}})$, raising the question of whether the learned representation can support stable evolution over physically relevant timescales. To address this, we perform ensemble rollout tests extending up to five autocorrelation times and examine coherence, field statistics, and transport quantities.

\subsection{Rollout protocol}

The trained model is initialized from multiple independent starting times ($t_0$), uniformly sampled across the statistically stationary portion of the simulation. For each rollout length, an ensemble of 40--60 trajectories is generated, allowing robust statistical assessment of stability and variability. Rollouts are performed for up to 1400 prediction steps, corresponding to approximately $(5T_{\mathrm{ac}})$.

\subsection{Computational performance and speedup}
\begin{comment}
To quantify the computational advantage of GKFieldFlow, we compare inference
cost to the underlying CGYRO simulation in matched physical units ($a/c_s$)
using total GPU resource usage (GPU--seconds). For the DIII--D test case
considered here, advancing the nonlinear CGYRO model over $53\,a/c_s$
requires approximately $600$~s on $24$ NVIDIA A100 GPUs, corresponding to a
total cost of $\sim 1.4\times10^4$~GPU--seconds.

In contrast, GKFieldFlow produces an autoregressive rollout over the same
physical interval on a single GPU at a cost that depends on the temporal
window length. For the highest--accuracy configuration ($T_c=20$), inference
requires $\sim 58$~GPU--seconds. For compact temporal contexts that retain
excellent accuracy at $\Theta=6$ ($T_c=4$ and $T_c=8$), inference over one
turbulence autocorrelation time requires only $\sim 2.4$ and $\sim 4.2$
GPU--seconds, respectively. Using $1\,\tau_{\mathrm{auto}}\approx 21\,a/c_s$,
this corresponds to a CGYRO cost of $\sim 5.7\times10^3$~GPU--seconds per
autocorrelation time.

Thus, depending on the chosen temporal context, GKFieldFlow reduces total GPU
resource usage by up to $\mathcal{O}(10^3)$ relative to direct
nonlinear CGYRO evolution, while maintaining strong predictive fidelity. This
comparison isolates inference cost (excluding training) and highlights the
practical value of surrogate models for rapid exploration and real--time
applications.
\end{comment}
To quantify the computational benefit of the surrogate models we compare inference
costs to the underlying CGYRO simulation in matched physical units ($a/c_s$) for a range of temporal context lengths $T_c=1,4,8,20$, where $T_c=20$ is the optimal window with the lowest prediction errors. 

For the DIII--D test case considered here, advancing the nonlinear CGYRO
model over $53\,a/c_s$ requires approximately $600$\,s of wall--clock time
on 6 nodes with 4 GPUs per node (24 NVIDIA A100 GPUs on Perlmutter @ NERSC),
corresponding to a total cost of $\sim 1.4\times10^4$ GPU--seconds. In
contrast, GKFieldFlow produces an autoregressive rollout over the same
physical interval on a single GPU, with inference cost determined by $T_c$.
For the highest--accuracy configuration ($T_c=20$), inference over
$53\,a/c_s$ requires $\sim 58$ GPU--seconds, yielding a compute--normalized
speedup of approximately $2.5\times10^2$ relative to CGYRO.

Using the turbulence autocorrelation time
$\tau_{\mathrm{auto}}=9\,a/c_s$ as a natural physical unit, the CGYRO cost
corresponds to $\sim 2.4\times10^3$ GPU--seconds per autocorrelation time.
For compact temporal contexts that retain excellent accuracy at
$\Theta=6$, GKFieldFlow requires only $\sim 2.4$ and $\sim 4.2$ GPU--seconds
per autocorrelation time for $T_c=4$ and $T_c=8$, respectively. Scaling
these costs to the same $53\,a/c_s$ interval yields total surrogate costs of
$\sim 14$ and $\sim 25$ GPU--seconds, corresponding to speedups of
$\sim 1.0\times10^3$ and $\sim 5.7\times10^2$ relative to CGYRO.

For reference, the purely instantaneous baseline ($T_c=1$), which contains
no explicit temporal information, requires only $\sim 0.9$ GPU--seconds per
autocorrelation time. Over $53\,a/c_s$, this corresponds to a total cost of
$\sim 5$ GPU--seconds and a maximum compute--normalized speedup of
$\sim 2.6\times10^3$. This configuration provides a lower bound on
surrogate inference cost and illustrates the full extent of achievable
computational acceleration.

Taken together, these results demonstrate that GKFieldFlow offers a
controllable accuracy--cost envelope: higher $T_c$ yields improved accuracy
at increased cost, while compact temporal contexts ($T_c=4$--$8$) provide an
excellent balance between fidelity and efficiency. Depending on the chosen
operating point, the surrogate reduces total GPU resource usage by factors
ranging from $\mathcal{O}(10^2)$ to $\mathcal{O}(10^3)$ relative to direct
nonlinear gyrokinetic simulation, highlighting its value for rapid
exploration, parameter scans, and real--time applications.

\subsection{Phase coherence across toroidal modes}
Figure~\ref{fig:coherence} presents the circular phase coherence of the predicted electrostatic potential as a function of toroidal wavenumber $k_y$ and rollout duration.

\begin{figure}[t]
\centering

\centering
\includegraphics[width=\linewidth]{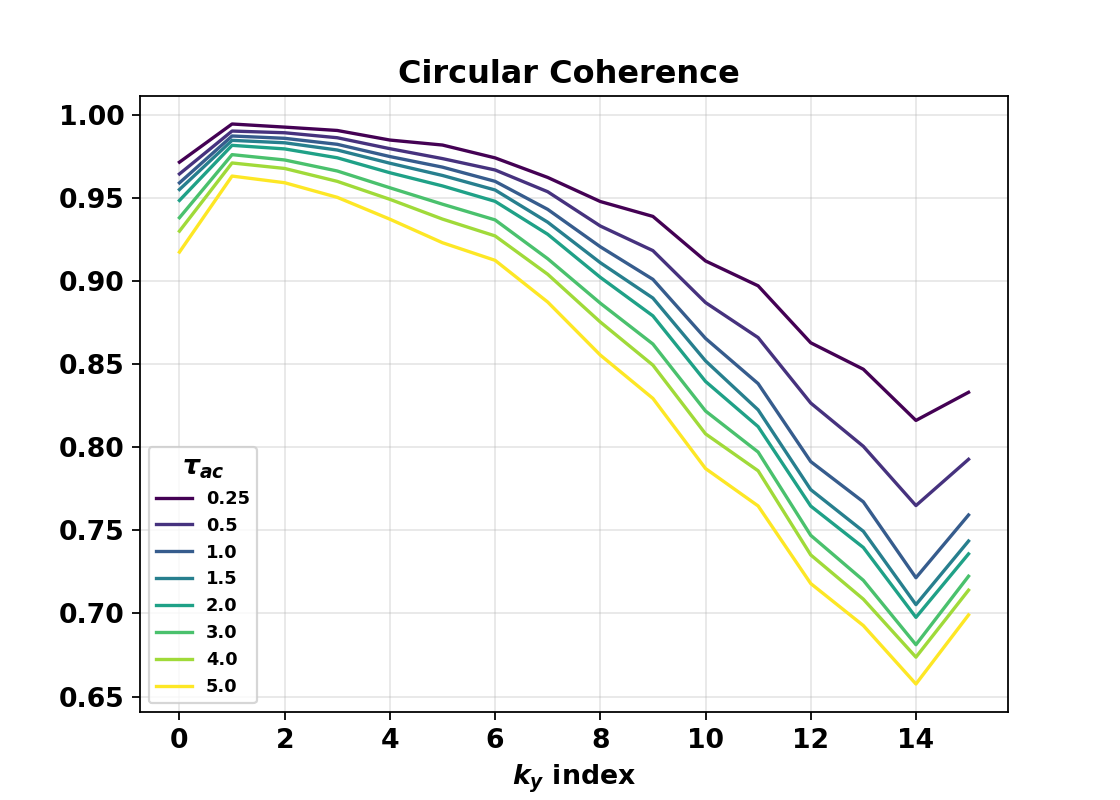}
\caption{Circular coherence versus $k_y$ for increasing
rollout duration in units of $T_{ac}$.}
\label{fig:coherence}
\end{figure}

\begin{figure}[t]
\centering
\includegraphics[width=\linewidth]{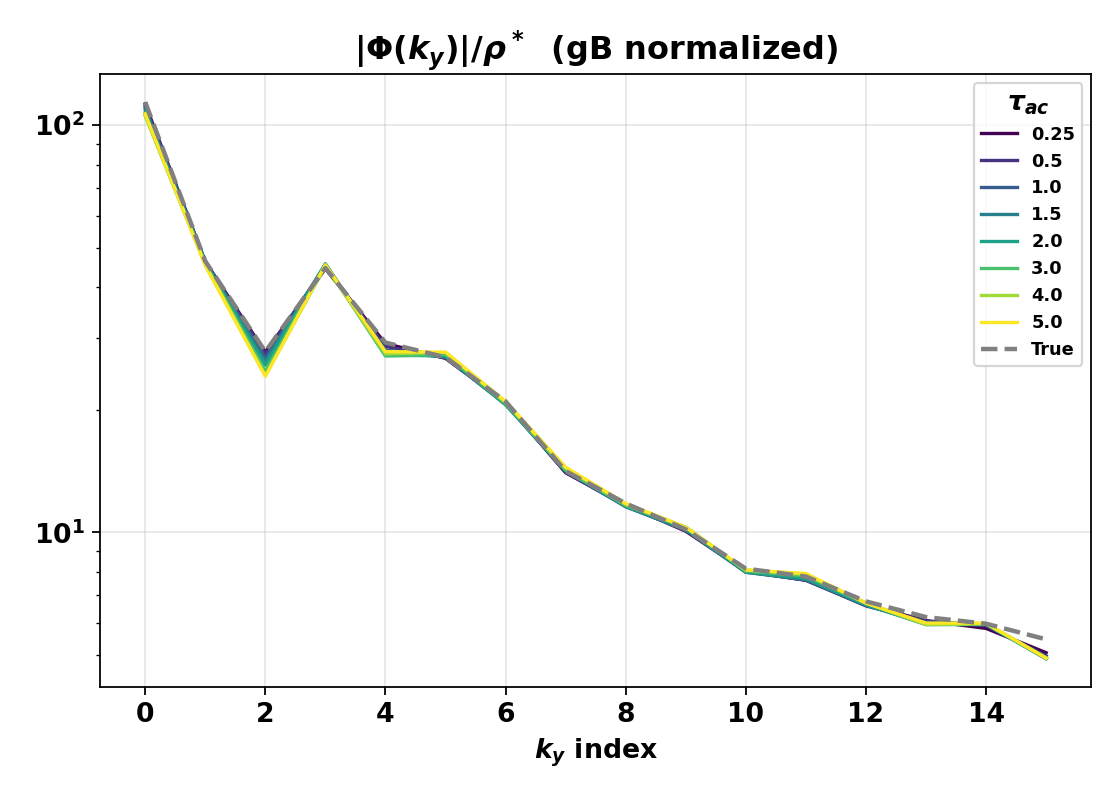}

\caption{Normalized amplitude distribution as a function of $k_y$ for increasing
rollout duration in units of $T_{ac}$.}
\label{fig:dist}
\end{figure}

Several key observations emerge. First, low-$k_y$ modes, including the zonal-flow component, remain highly coherent throughout the entire rollout window. Second, intermediate $k_y$ modes retain substantial coherence even beyond one autocorrelation time, with a gradual and monotonic decay rather than abrupt decorrelation. Third, high-$k_y$ modes decorrelate more rapidly, but still maintain nontrivial coherence levels even at $5T_{\mathrm{ac}}$. This hierarchy is consistent with gyrokinetic physics, where large-scale energy-containing modes dominate transport and exhibit longer memory than small-scale fluctuations.

GyroBohm normalized amplitude distribution versus toroidal wave-number is shown in Fig.~\ref{fig:dist} for all the rollout durations. Overall, amplitude distribution stays very close to the true value. At $5T_{\mathrm{ac}}$ the maximum relative deviation from the true ampltides is at $k_y(2)\approx 13\%$.
Together Figs.~\ref{fig:coherence} and~\ref{fig:dist} show that the \GKFieldFlow is capable of predicting the  turbulence fields phases and amplitudes to high accuracy for individual $k_y$ modes in long-horizon autoregressive predictions. 

\subsection{Potential amplitude stability under rollouts}

Figure~\ref{fig:potential_rollout} shows the normalized root-mean-square error (NRMSE) of the potential amplitude as a function of rollout duration, separated into zonal-flow $(k_y = 0)$ and turbulent $(k_y > 0)$ contributions.

\begin{figure}[t]
\centering
\begin{subfigure}{0.85\linewidth}
\centering
\includegraphics[width=\linewidth]{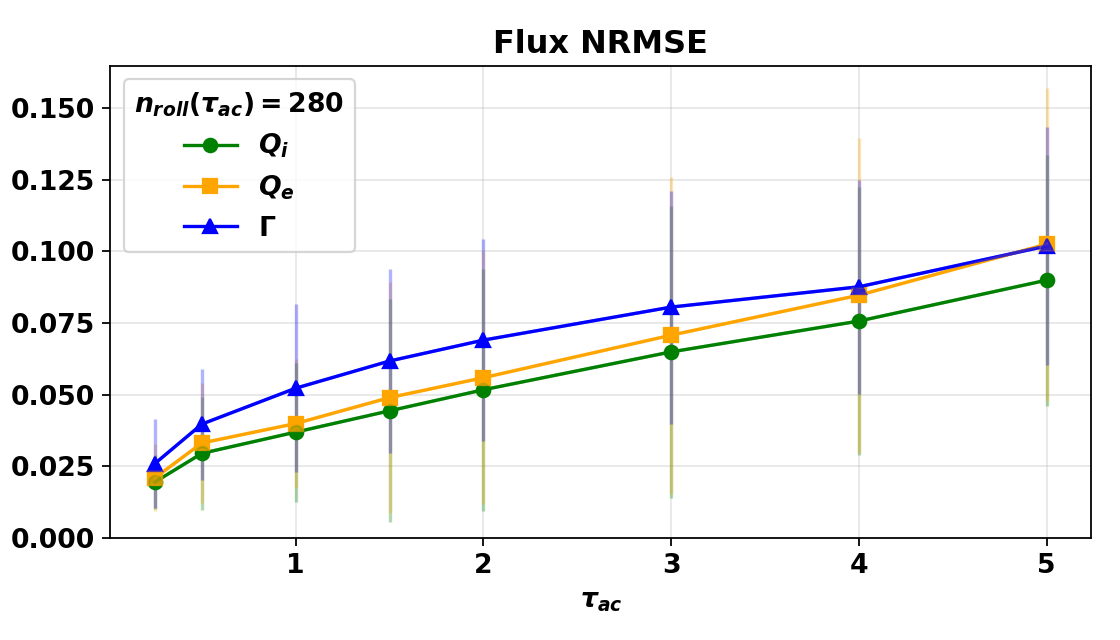}
\caption{\label{fig:flux_rollout}}
\end{subfigure}
\hfill
\begin{subfigure}{0.85\linewidth}
\centering
\includegraphics[width=\linewidth]{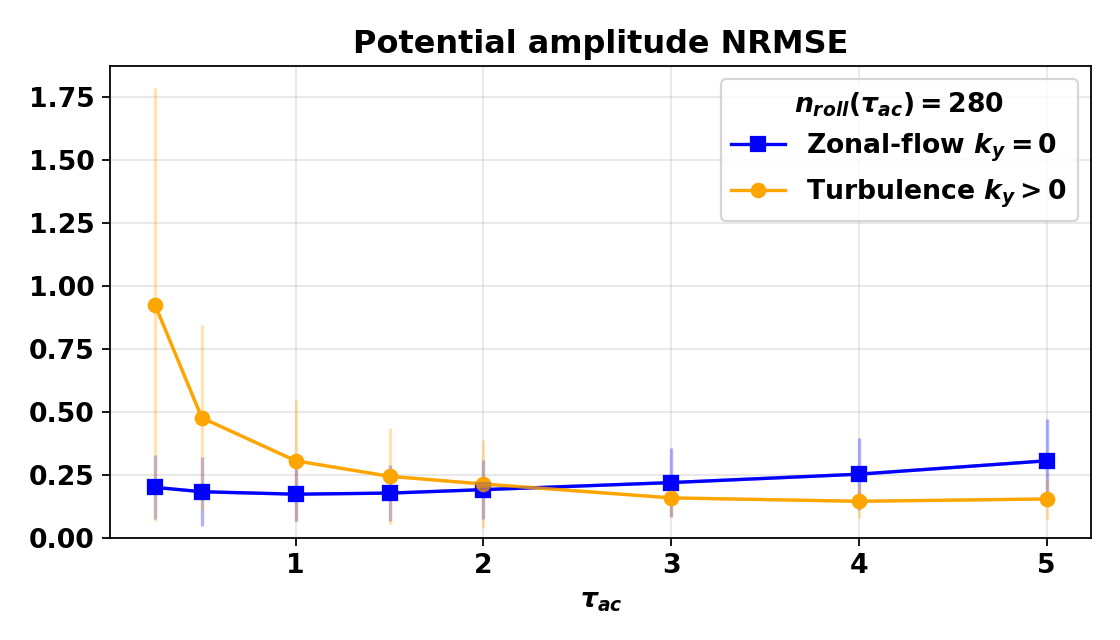}
\caption{\label{fig:potential_rollout}}
\end{subfigure}
\caption{(a) Flux NRMSE versus rollout duration for $Q_i$, $Q_e$, and $\Gamma$. (b) Potential amplitude NRMSE versus rollout duration for zonal-flow and turbulent components.}

\end{figure}

The turbulent component exhibits low initial error and only a slow increase with rollout length, saturating at modest values even at $5T_{\mathrm{ac}}$. The zonal-flow component shows a slightly stronger upward trend, reflecting its longer intrinsic memory and sensitivity to cumulative phase error. Crucially, neither component exhibits runaway growth or exponential error amplification, indicating that the learned dynamics remain well conditioned under autonomous evolution.

\subsection{Flux prediction accuracy over long horizons}

Figure~\ref{fig:flux_rollout} presents the NRMSE of ion heat flux $(Q_i)$, electron heat flux $(Q_e)$, and particle flux $(\Gamma)$ as a function of rollout duration.

All three transport channels exhibit smooth, gradual error growth with increasing rollout length, remaining well bounded even at five autocorrelation times. No sharp transitions or instabilities are observed. This behavior demonstrates that accurate transport prediction does not require resolving the full turbulence autocorrelation time during training, and that the surrogate captures the dominant transport-relevant dynamics.

\subsection{Time-resolved rollout trajectories}

Representative time traces of predicted and true quantities are shown in Fig.~\ref{fig:timeseries} and ~\ref{fig:timeseries_amps}, comparing rollouts of one and five autocorrelation times. Shown are the heat and particle fluxes, the turbulent fluctuation intensity $(k_y > 0)$, and the zonal-flow intensity $(k_y = 0)$.

\begin{figure}[t]
\centering
\begin{subfigure}{0.85\linewidth}
\includegraphics[width=\linewidth]{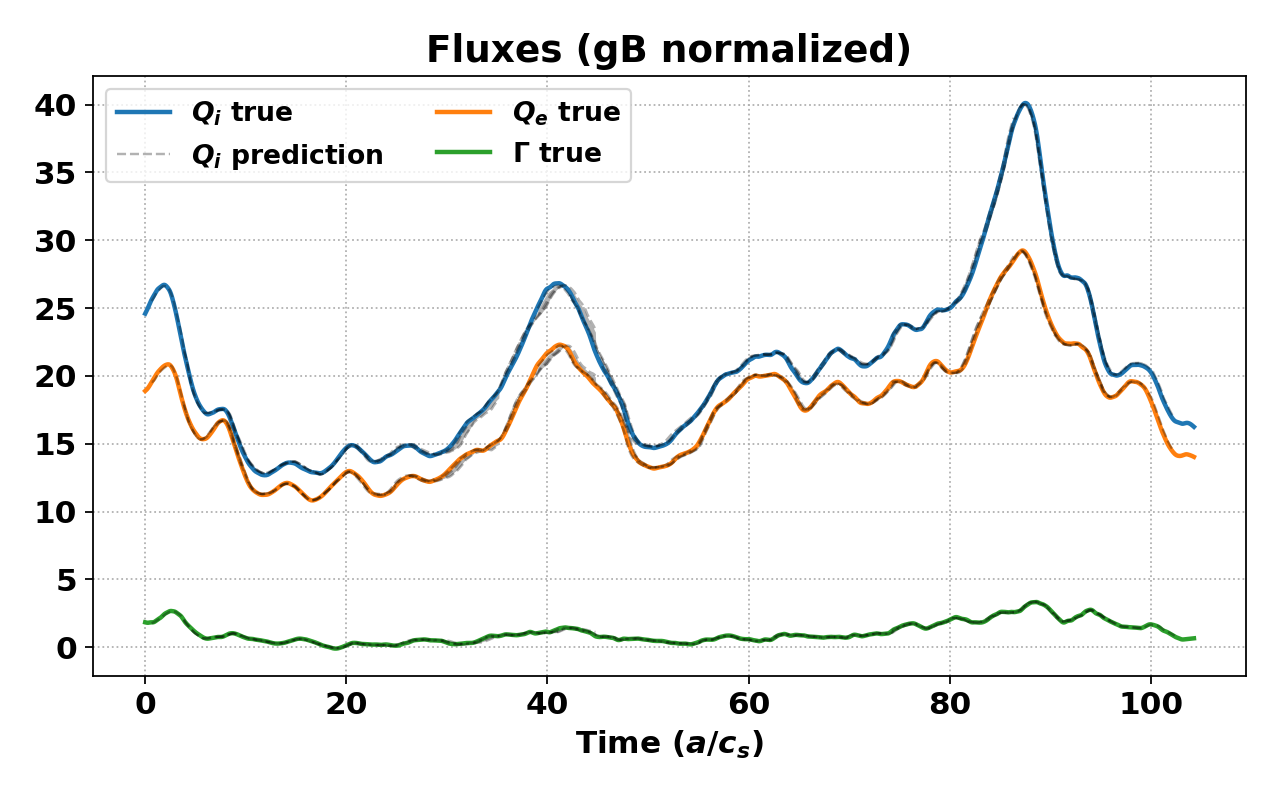}
\caption{$t_{\text{roll}} =T_{\mathrm{ac}}  $ } 
\end{subfigure}
\begin{subfigure}{0.85\linewidth}
\includegraphics[width=\linewidth]
{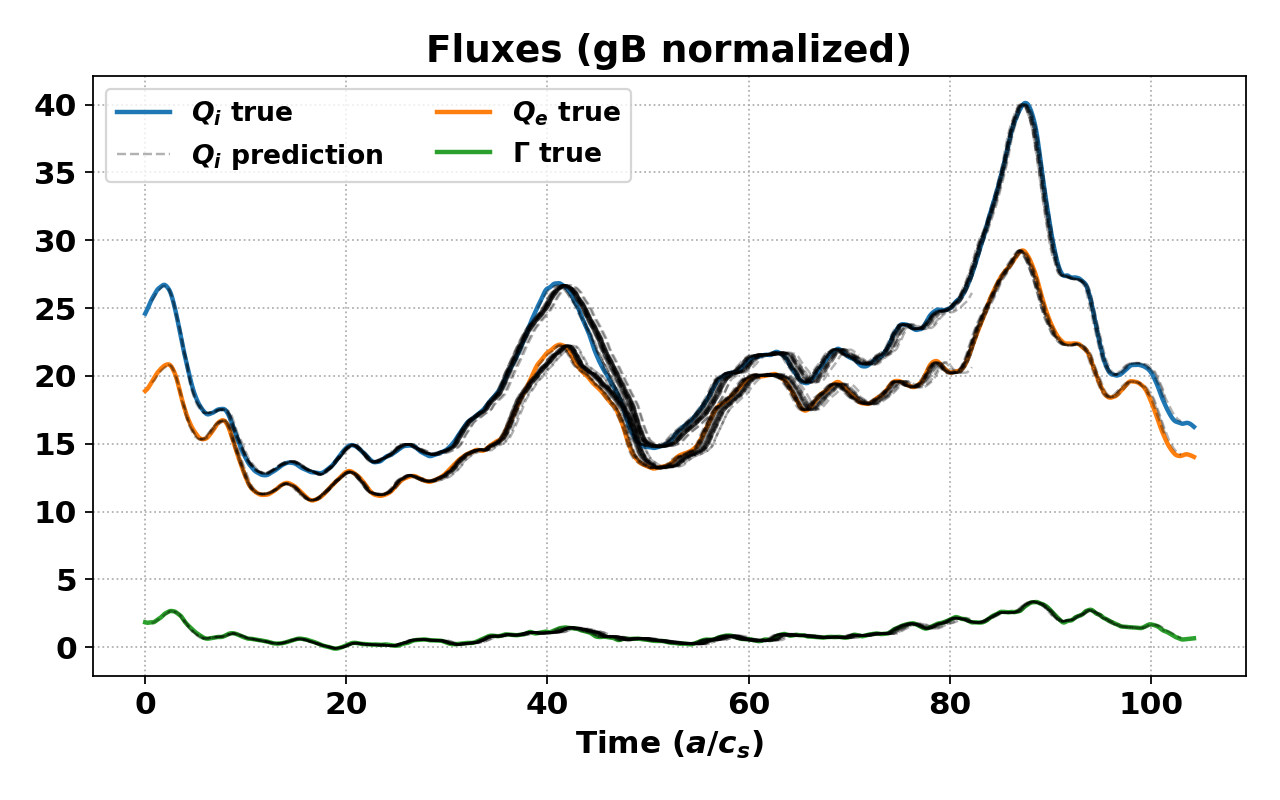}
\caption{$t_{\text{roll}} =5T_{\mathrm{ac}}  $ } 
\end{subfigure}
\caption{Representative flux rollout time series comparing true and predicted quantities for $1T_{\mathrm{ac}}$ (top) and $5T_{\mathrm{ac}}$ (bottom). $T_{\mathrm{ac}}$ is the turbulence autocrrelation time $\sim 9 a/c_s$.}
\label{fig:timeseries}
\end{figure}

\begin{figure}[t]
\centering
\begin{subfigure}{0.85\linewidth}
\includegraphics[width=\linewidth]{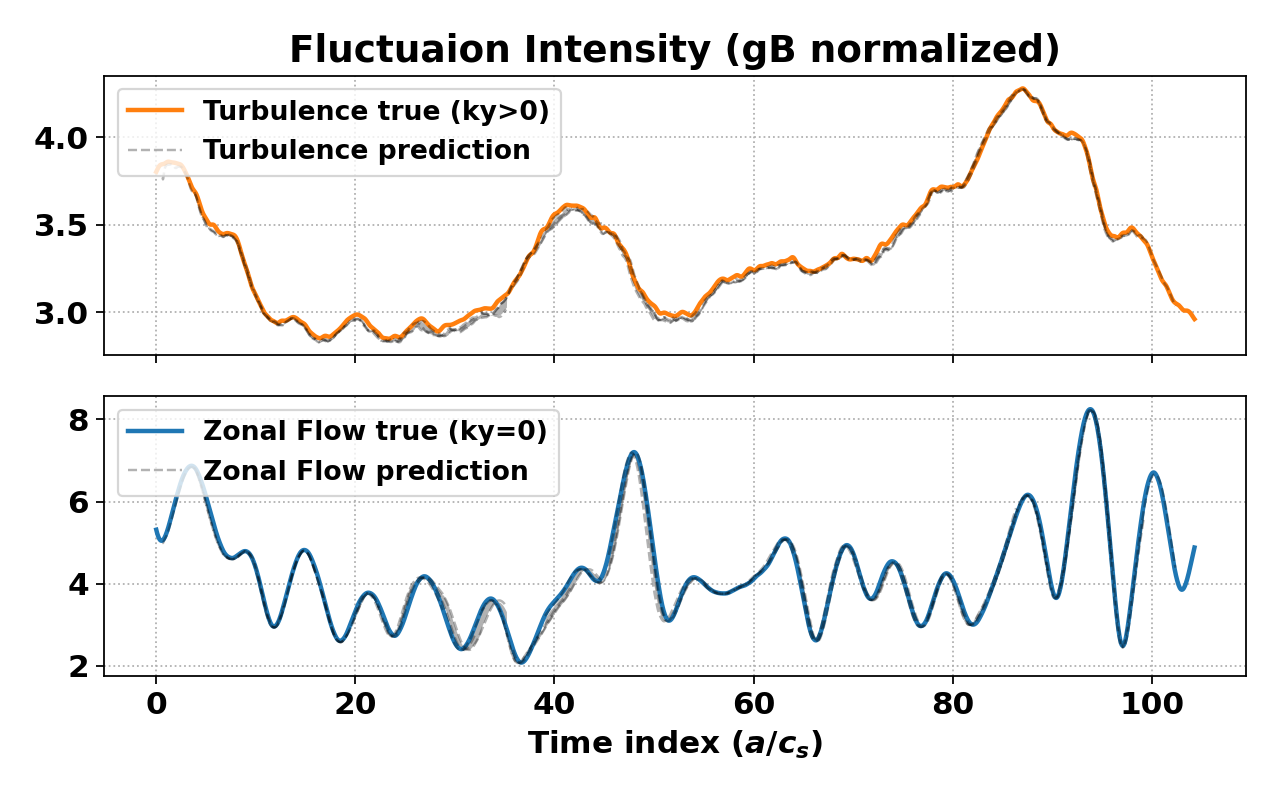}
\caption{$t_{\text{roll}} =T_{\mathrm{ac}}  $ } 
\end{subfigure}
\begin{subfigure}{0.85\linewidth}
\includegraphics[width=\linewidth]{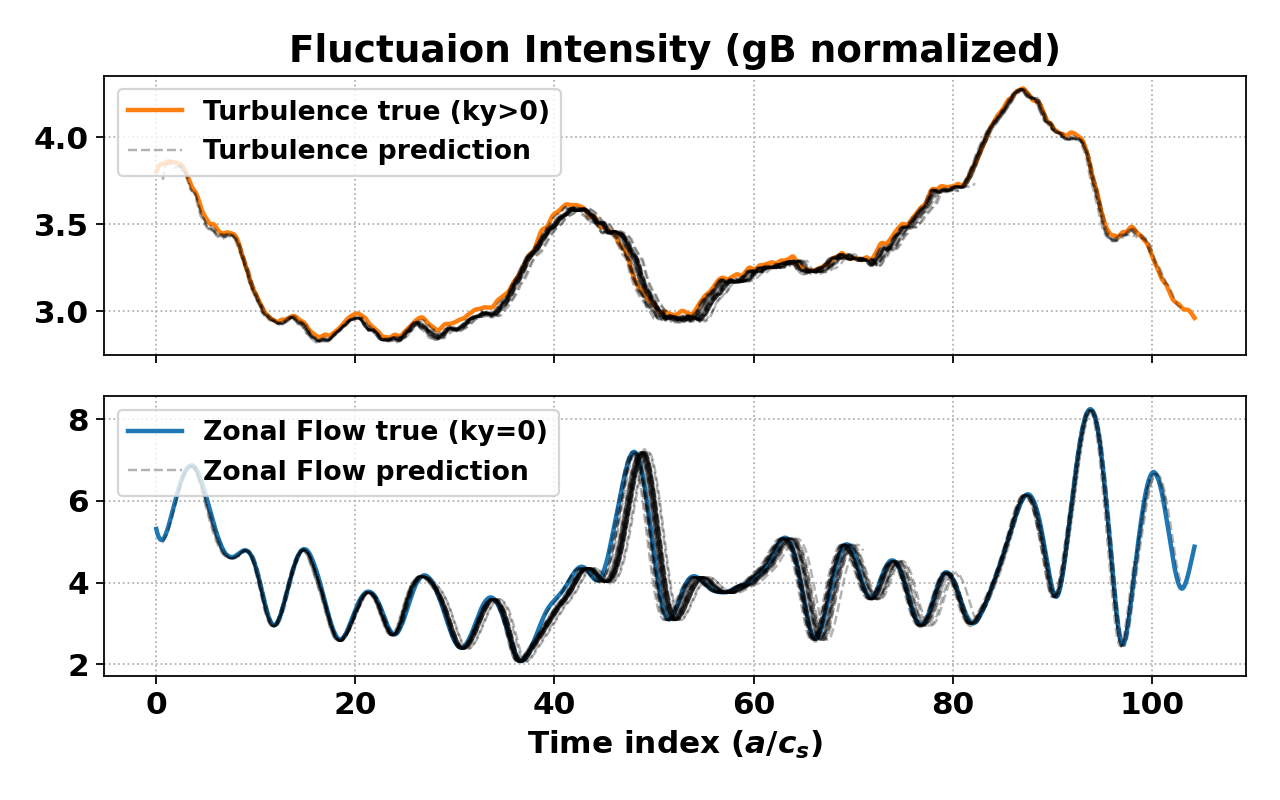}
\caption{$t_{\text{roll}} =5T_{\mathrm{ac}}  $ } 
\end{subfigure}
\caption{Representative turbulence and zonal flow amplitude rollout time series comparing true and predicted quantities for $1T_{\mathrm{ac}}$ (top) and $5T_{\mathrm{ac}}$ (bottom).}
\label{fig:timeseries_amps}
\end{figure}
At one autocorrelation time, predictions are nearly indistinguishable from the ground truth. Remarkably, even at five autocorrelation times, the surrogate reproduces peak timing, amplitudes, and long-time envelopes with high fidelity. Ensemble trajectories remain tightly clustered, indicating robust and stable behavior. Note that the model accurately reproduces: (i) the amplitude and timing of peak flux events, (ii) the intermittent bursting character of the turbulence intensity, and (iii) the large-amplitude oscillations in the zonal flow channel. This indicates that the surrogate has learned the physically correct correlation between field dynamics and transport rather than simply fitting mean values.

\subsection{Physical interpretation and implications}

Taken together, these results demonstrate that the \GKFieldFlow surrogate learns a compact, task-relevant representation of the turbulent state rather than memorizing explicit long-time correlations. Although the optimal training context $T_c = 20$ is far shorter than the turbulence autocorrelation time, the inferred latent state supports stable autonomous evolution over many $T_{\mathrm{ac}}$. The observed scale-dependent coherence and transport accuracy further indicate that the model has learned the correct hierarchy of temporal stiffness across modes, consistent with gyrokinetic physics.

\subsection{Conclusion}

The long-horizon rollout tests presented here establish that the \GKFieldFlow surrogate trained at the optimal context length $T_c = 20$ is both stable and physically faithful under free-running evolution. Accurate phase coherence, bounded field errors, and robust transport prediction persist for rollouts extending up to five turbulence autocorrelation times. These results validate the interpretation of a finite effective learned memory that is substantially shorter than the physical autocorrelation time, while still enabling long-horizon predictive capability. This property is essential for future applications of field-based surrogates to reduced modeling, control, and fast transport prediction in gyrokinetic turbulence.
\begin{figure}[t]
\centering
\begin{subfigure}{1.\linewidth}
\includegraphics[width=\linewidth]{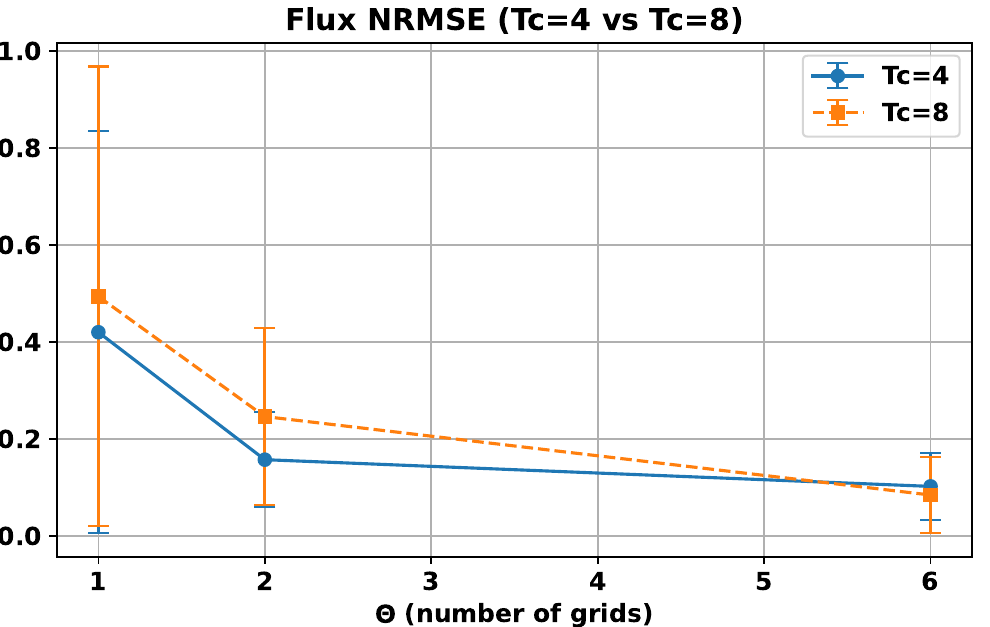}
\caption{ } 
\end{subfigure}
\begin{subfigure}{1.0\linewidth}
\includegraphics[width=\linewidth]
{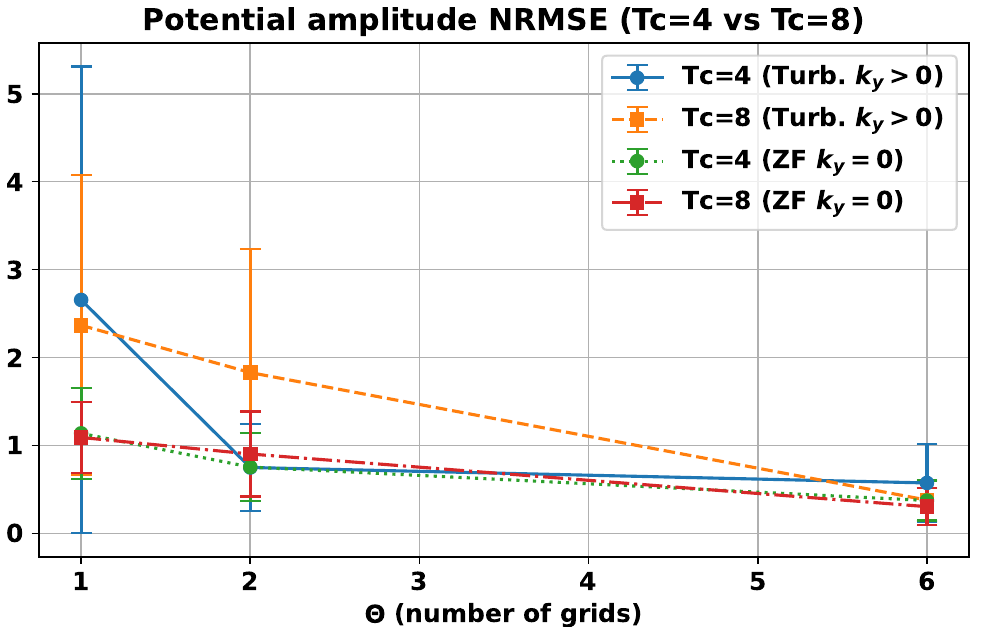}
\caption{ } 

\end{subfigure}
\caption{(a) Normalized RMSE of (a) Fluxes (b) turbulence and zonal flow, versus the number of poloidal grid points ($\Theta$) used for training \GKFieldFlow. RMSE are from rollouts with 280 steps, equivalent to one turbulence autocorrelation time ($\sim 9 a/c_s$). For both time contexts $T_c=4,8$, reducing $\Theta$ results in larger errors.}
\label{fig:theta_ablation_rmse}
\end{figure}
\begin{comment}
\begin{figure}[t]
\centering
\begin{subfigure}{0.85\linewidth}
\includegraphics[width=\linewidth]{spaghetti_phi_envelope.png}
\caption{$t_{\text{roll}} =T_{\mathrm{ac}}  $ } 
\end{subfigure}
\begin{subfigure}{0.85\linewidth}
\includegraphics[width=\linewidth]{spaghetti_amp_envelope_5.png}
\caption{$t_{\text{roll}} =5T_{\mathrm{ac}}  $ } 
\end{subfigure}
\caption{Representative turbulence and zonal flow amplitude rollout time series comparing true and predicted quantities for $1T_{\mathrm{ac}}$ (top) and $5T_{\mathrm{ac}}$ (bottom).}
\label{fig:timeseries_amps}
\end{figure}
\end{comment}

\begin{figure}[t]
\centering
\begin{subfigure}{0.9\linewidth}
\includegraphics[width=\linewidth]{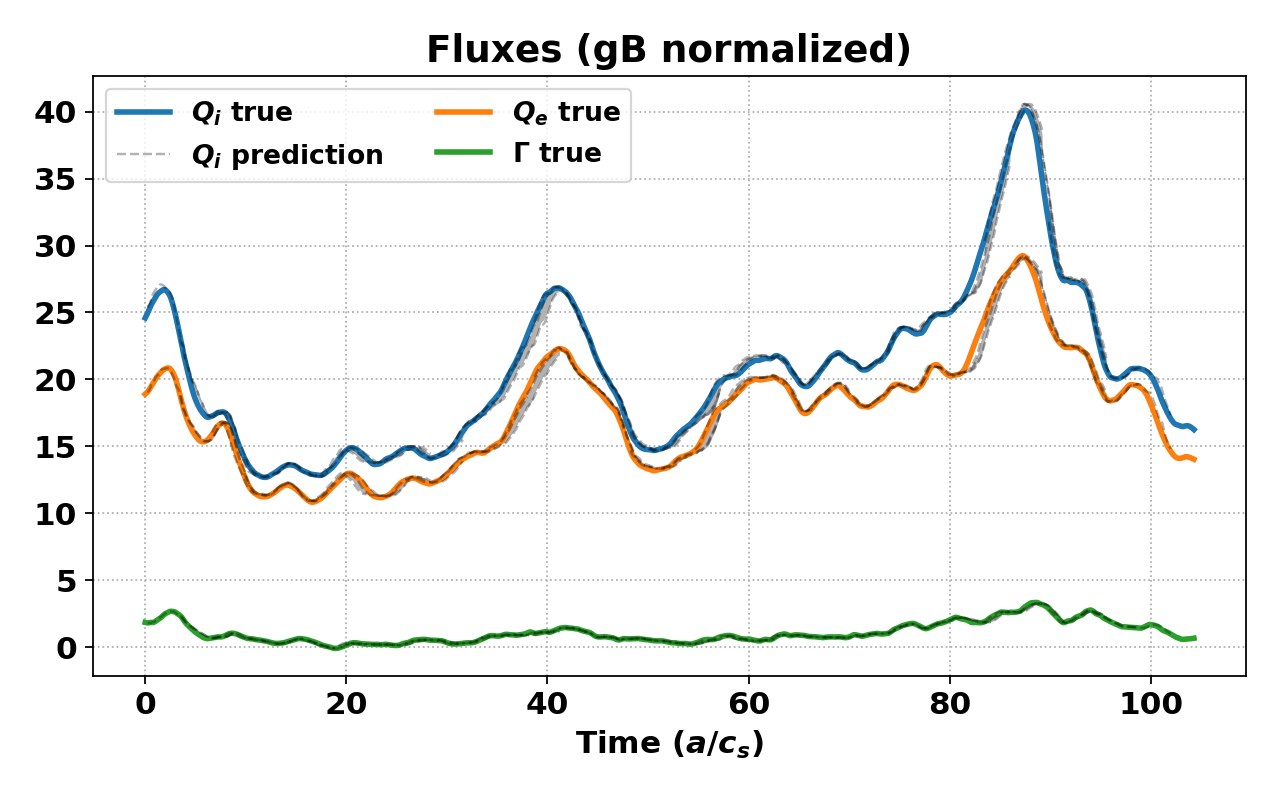}
\caption{ } 
\end{subfigure}
\begin{subfigure}{0.9\linewidth}
\includegraphics[width=\linewidth]{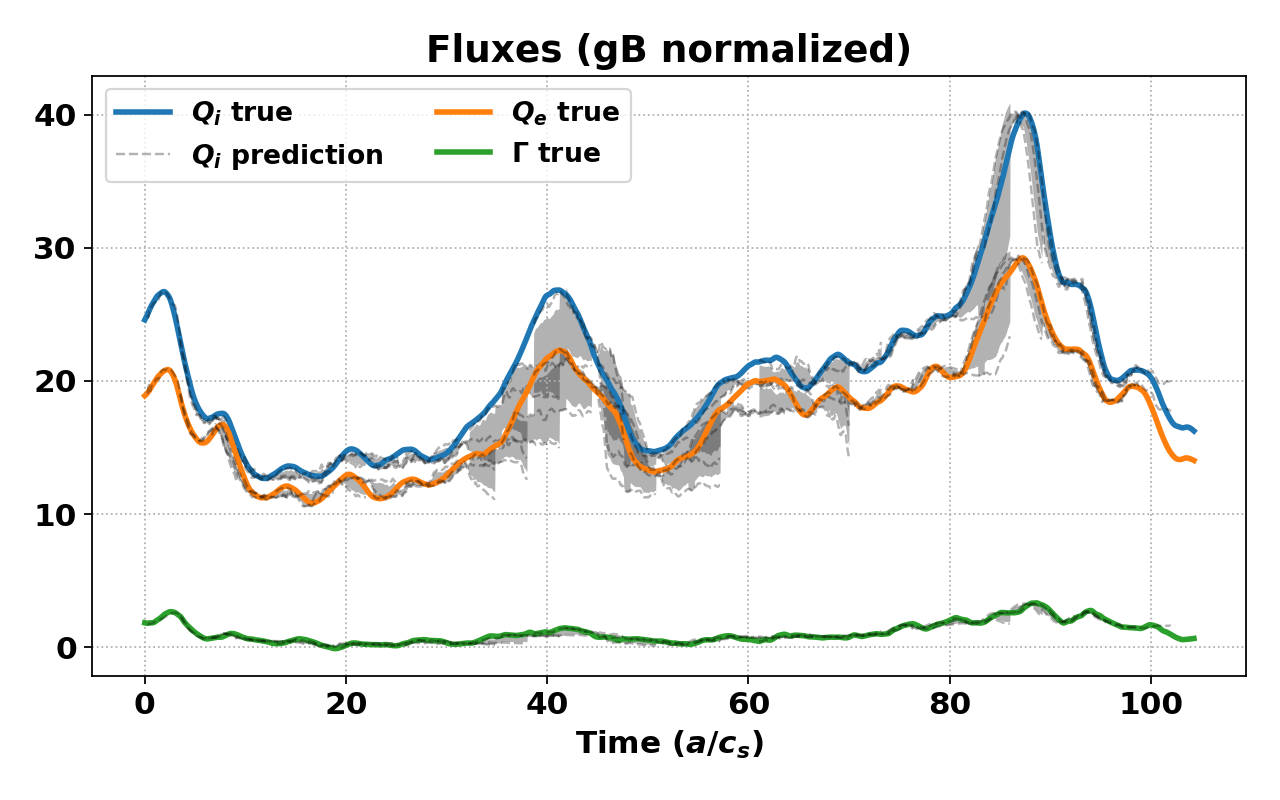}
\caption{ } 
\end{subfigure}
\begin{subfigure}{0.9\linewidth}
\includegraphics[width=\linewidth]{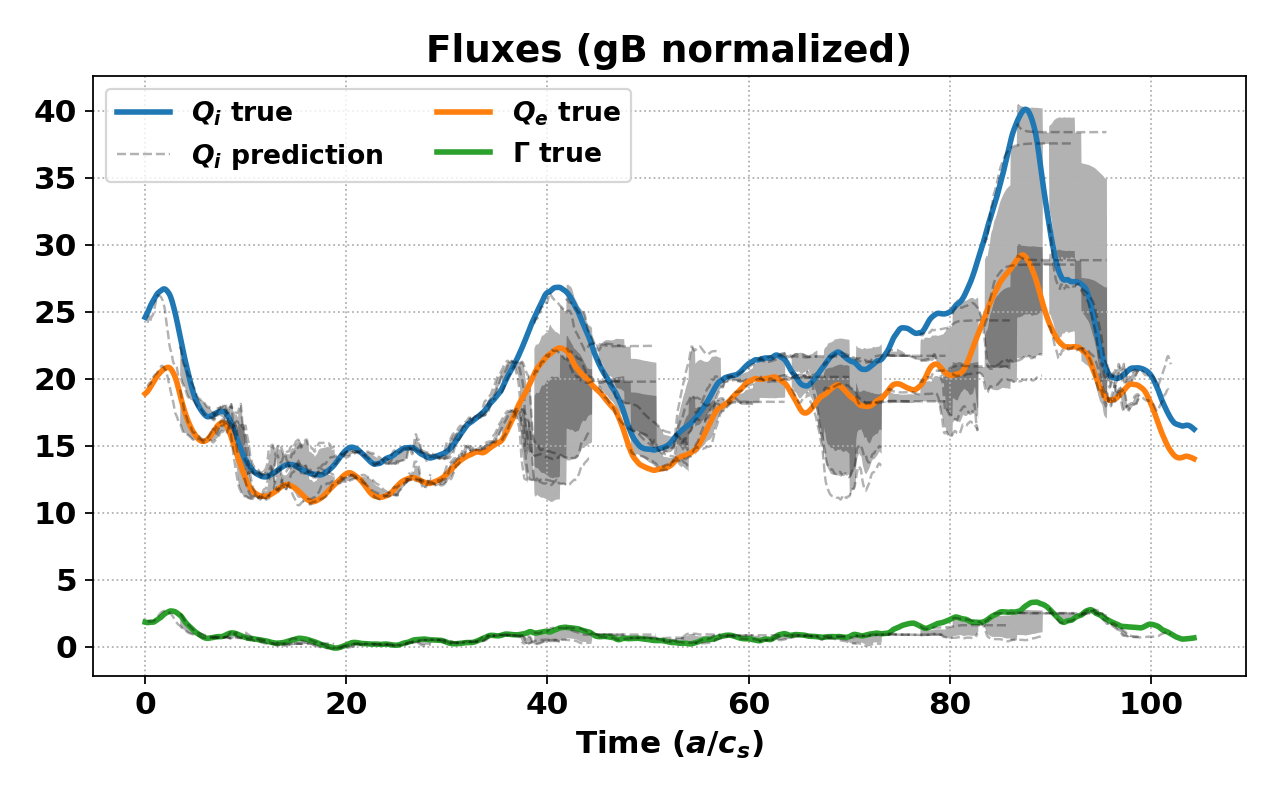}
\caption{ } 
\end{subfigure}
\caption{Spaghetti plots of an ensemble of rollouts for $T_c=8$ with uniformly distributed initial times, for $\Theta=6,2,1$ respectively in (a),(b and (c). RMSE are from rollouts with 280 steps, equivalent to one turbulence autocorrelation time ($\sim 9 a/c_s$).}
\label{fig:theta_spaghetti_rollouts}
\end{figure}

\section{Sensitivity to poloidal resolution ($\Theta$) and information sufficiency}

To assess the role of the poloidal structure in constraining the accuracy of the surrogate and
the stability of the long-horizon, we performed a targeted ablation study in which the
number of retained poloidal grid points $\Theta$ was systematically reduced.
This experiment was designed as a stress test: by progressively removing
poloidal information, we intentionally degrade the physical content of the input
and examine how both single-time inference accuracy and autoregressive rollout
behavior respond.

Figure~\ref{fig:theta_ablation_rmse} summarizes the resulting errors for
$\Theta=\{1,2,6\}$ at representative temporal window lengths ($T_c=4$ and
$T_c=8$). Normalized RMSE are from rollouts with 280 steps, equivalent to one turbulence autocorrelation time ($\sim 9 a/c_s$). $\Theta=2$ keeps the outboard and inboard midplane grids of the tokamak palsma cross section and $\Theta=1$ only keeps the outboard midplane grid.
A clear and monotonic trend emerges. Models trained with $\Theta=6$
exhibit consistently low flux RMSE and substantially reduced errors in both
turbulence and zonal-flow amplitudes. Reducing the poloidal resolution to
$\Theta=2$ leads to a marked degradation across all metrics, while the extreme
outboard-midplane-only case ($\Theta=1$) produces large errors and significantly
increased variance. This behavior is consistent with the loss of essential
poloidal structure, including ballooning localization and phase relationships,
that are known to be critical for determining transport and zonal-flow dynamics
in gyrokinetic turbulence.

The impact of reduced poloidal information is better perceived from the plots
of autoregressive rollout versus time. Figure~\ref{fig:theta_spaghetti_rollouts} shows
representative ``spaghetti'' plots of flux evolution for $\Theta=1$, $\Theta=2$,
and $\Theta=6$. For $\Theta=6$, the surrogate maintains tight trajectory
bundling, accurate mean behavior, and stable long-horizon evolution over
multiple turbulence autocorrelation times. At $\Theta=2$, the rollouts exhibit
noticeably increased spread and intermittent deviations, indicating reduced
ability to remain on the physically admissible manifold. In contrast, the
$\Theta=1$ case shows large trajectory divergence and rapid loss of fidelity,
demonstrating that a single poloidal slice is insufficient to constrain the
nonlinear dynamics over long horizons.

Together, these results demonstrate that poloidal resolution plays a central role in enabling high-fidelity surrogate modeling of gyrokinetic turbulence. The controlled degradation observed as $\Theta$ is reduced provides strong evidence that GKFieldFlow leverages physically meaningful poloidal structure—including ballooning localization, phase relationships, and perpendicular mode coupling—to learn a stable representation of the turbulent attractor. This ablation study validates that the model's success is not due to exploiting simple statistical correlations, but rather depends on capturing the essential physics encoded in the full three-dimensional field structure.

\section{\label{summ}Discussion and Future Work}

The results presented in this work demonstrate that the spatio--temporal neural architecture of \GKFieldFlow which combines a three-dimensional U-Net with a temporal model can accurately predict turbulent transport from saturated gyrokinetic field data. The present study focuses on a representative ion-scale case to establish feasibility. Systematic evaluation across parameter space (varying geometry, collisionality, $\beta$, etc.) and electromagnetic regimes will be the subject of forthcoming work. In practical terms, for this test case the surrogate provides $\mathcal{O}(10^3)$ reduction in GPU resources relative to CGYRO for the same physical interval, while maintaining coherent field structure and accurate flux statistics.

Beyond its performance for the specific test-case presented here, the model design is intentionally modular and admits several natural extensions, which we outline below.

\subsection{Extension to kinetic moments and additional gyrokinetic fields}

In the present study, the fluctuating electrostatic potential $\phi(\mathbf{r},t)$ serves as the sole field-level input, represented by its real and imaginary components as separate channels. However, the proposed architecture is readily generalizable to incorporate additional gyrokinetic outputs that are routinely computed by nonlinear GK simulations. In particular, kinetic moments such as density, temperature, and parallel flow fluctuations ($\delta n_s$, $\delta T_s$, $\delta u_{\parallel,s}$), as well as electromagnetic fields such as $A_\parallel$ (and $\delta B_\parallel$ where relevant), are likewise complex-valued, three-dimensional quantities defined on compatible spatio--temporal grids.

These observables can therefore be introduced as additional input channels without modification to the core architecture. From a physical standpoint, this flexibility enables systematic exploration of how access to progressively richer phase-space information affects transport prediction, robustness, and generalization. In particular, the inclusion of kinetic moments may provide the model with more direct proxies for energy and particle transport pathways, potentially improving extrapolation behavior.

\subsection{Spectral locality and implications for multiscale turbulence}

The simulation examined here corresponds to ion-scale turbulence in which nonlinear coupling is relatively local in $k_y$ space, including interactions between zonal flows and the dominant turbulent modes. As a consequence, accurate transport prediction is achieved with a modest effective receptive field depth, and no explicit downsampling is required in the binormal ($k_y$) direction.

This situation is not expected to be generic. In multiscale ion--electron turbulence, as well as in ETG-dominated regimes, the active spectral bandwidth broadens substantially and cross-scale interactions become more prominent. Efficient representation of such dynamics may require architectural extensions that explicitly encode multiresolution structure in $k_y$, such as downsampling--upsampling pathways or hierarchical spectral representations. Importantly, such approaches must preserve the fidelity of low-$k_y$ components, which often play a disproportionate role in regulating transport through zonal-flow dynamics.

These considerations suggest that architectural choices optimal for single-scale ion turbulence may not transfer directly to multiscale regimes, and that the present framework provides a flexible starting point for investigating these trade-offs in a controlled manner.
\begin{table*}[t]
\centering
\renewcommand{\arraystretch}{1.15}
\setlength{\tabcolsep}{5pt}
\scriptsize
\begin{tabular}{l l l l}
\hline
\textbf{Domain} &
\textbf{Input Channels $\mathbf{X}$} &
\textbf{Predicted Targets $\mathbf{y}$} &
\textbf{Use-Case / Interpretation} \\
\hline
\textbf{Gyrokinetic (Fusion)} &
$\{\phi,\;A_\parallel,\;\delta n,\;\dots\}$ &
$\{Q_i,\;Q_e,\;\Gamma\}$ &
Turbulent transport surrogate; \\
&&&$\mathcal{O}(10^3)$ faster flux evaluation. \\

\textbf{Fluid / CFD / Aero} &
$\{\mathbf{u},\;\omega,\;p\}$ &
$\{\tau_{\mathrm{SGS}},\;\epsilon,\;\Pi\}$ &
Subgrid closures, dissipation estimates,\\
&&&reduced-order modeling. \\

\textbf{Climate / GCM Physics} &
$\{T,\;q,\;\mathbf{u},\;\partial_z \theta\}$ &
Heat/cooling, moisture tendencies &
Rapid parametrization \\
&&&replacement for GCM physics. \\

\textbf{Waves / EM / Photonics} &
$\{\mathbf{E},\mathbf{B},\psi\}$ (real/complex) &
Envelope/radiation source terms &
Surrogates for dispersive operators\\
&&&and energy transfer rates. \\

\textbf{General PDE Surrogates} &
$\mathbf{X}(t\!-\!T_c\!:\!t)$ arbitrary channels &
$\mathcal{F}[\mathbf{X}]$ operator-defined outputs &
Template applies when locality \\
&&&+ causal evolution are present. \\
\hline
\end{tabular}
\caption{\label{tab:domain_mapping} Domain-agnostic mapping of the spatio-temporal surrogate \FieldFlowNet architecture. \GKFieldFlow is the gyrokinetic realization of a general template—3D spatial encoding, temporal modeling, and a domain-specific prediction head—while the input channels and physical operators vary by application.}
\end{table*}
\subsection{Toward a conditional local gyrokinetic surrogate}

A key long-term objective is the development of a general local gyrokinetic neural surrogate capable of interpolating across a broad parameter space of plasma conditions. After identifying an appropriate set of fluctuating-field and moment channels, the present model can be trained on a large database of CGYRO simulations spanning equilibrium geometry and drive parameters, including magnetic shaping, safety factor, magnetic shear, and normalized density and temperature gradients.

In this setting, simulation descriptors can be provided to the network as additional conditioning inputs, enabling a single model to represent a parametric family of local turbulence states. Such a conditional surrogate would move beyond single-regime emulation toward a data-driven approximation of the local gyrokinetic transport mapping itself. Rigorous assessment of generalization would require holding out entire regions of parameter space during training, thereby quantifying interpolation and limited extrapolation capability.

\subsection{Limitations}

Several limitations of the present study should be emphasized. First, the model is trained and evaluated on an individual set of ion-scale simulations with fixed physics assumptions and a limited range of parameters. Its performance outside this distribution, including across different geometries or electromagnetic regimes, has not been assessed. Second, the inputs are drawn from statistically saturated turbulence; the ability of the model to handle transient phases or regime transitions remains an open question. Finally, while the model captures correlations relevant for transport prediction, it does not enforce conservation laws explicitly, and its predictions should therefore be interpreted as data-driven approximations rather than reduced physical models.

Addressing these limitations—through broader training datasets, richer physics inputs, and incorporation of physically motivated inductive biases—represents a clear direction for future work.

\subsection{Outlook}

Taken together, these extensions outline a path toward neural surrogate models that operate directly on high-dimensional gyrokinetic field data while remaining closely aligned with the underlying physics. By systematically increasing physical richness, spectral complexity, and parametric coverage, the \GKFieldFlow framework provides a scalable platform for investigating the limits of data-driven transport modeling in gyrokinetic turbulence.

\subsection{Broader applicability of the architecture}

Although developed and demonstrated here for gyrokinetic turbulence, the proposed spatio--temporal \FieldFlowNet architecture is not specific to gyrokinetic turbulence, but represents a reusable spatio–temporal surrogate pattern: a 3D spatial feature encoder,
a temporal module, and a lightweight domain adapter that maps internal latent
representations to physically meaningful quantities. The same structure applies to a
broader class of PDE-governed systems in which (i) multi-channel fields evolve under
local temporal evolution and (ii) the desired outputs correspond to aggregate physical
operators rather than full field reconstructions. Table~\ref{tab:domain_mapping}
summarizes this portability. In this view, GKFieldFlow is a \textit{gyrokinetic
instantiation} of a general template; portability is achieved by changing only the
input channel semantics and output operators, rather than the architecture itself.

Examples include subgrid-scale modeling in fluid turbulence, where velocity or vorticity fields are mapped to stresses or dissipation rates; geophysical and climate modeling, where coarse-grained fluxes or tendencies depend on the recent evolution of atmospheric or oceanic fields; and wave-based systems involving complex-valued electromagnetic or acoustic fields, where phase information plays a central role. In all cases, the architecture operates as a data-driven functional mapping from spatio--temporal field data to physically meaningful aggregate quantities.

At the same time, successful application to other domains would require domain-specific choices of input channels, normalization, and training datasets, and no claim of universality is implied. Rather, the present work demonstrates a general modeling strategy for learning transport-relevant dynamics directly from field-level simulation data, which may be adapted to other physics contexts with similar structural characteristics.

\appendix
\section{\label{app:arch} DETAILED ARCHITECTURE SPECIFICATION}
\subsection{3D U-Net encoder for spatial feature extraction}

For each time slice $\Phi(t)$, the 3D spatial encoder applies a hierarchy of 3D convolutions using small cubic kernels ($n\times n \times n$) combined with multi-level spatial downsampling in the
$(R,\Theta)$ directions to extract spatial structures from turbulent fields. The feature channels increase with depth while the spatial resolution in the
$(R,\Theta)$ plane is progressively reduced. At each encoder level, strided
convolutions perform downsampling in $(R,\Theta)$, enabling the network to
capture increasingly global spatial structure while maintaining computational
tractability. The $k_y$ dimension is not downsampled and instead acts as an
independent spectral index carried through all encoder stages.

Each resolution level produces a corresponding multiscale feature embedding
$E_\ell$, which is later reused via skip connections in the decoder. Within
each level, residual 3D convolutional blocks with normalization and nonlinear
activation are employed to facilitate stable training and effective extraction
of multiscale geometric and spectral features from the turbulent fields.
\paragraph{Latent representation.}
At the coarsest spatial resolution, the encoded feature tensor is spatially
pooled and projected onto a compact latent vector,
\begin{equation}
    f_t \in \mathbb{R}^{d_{\mathrm{lat}}},
\end{equation}
which serves as a high-level representation of the instantaneous turbulent
state at time~$t$. The latent dimensionality $d_{\mathrm{lat}}$ is selected to
balance representational expressivity with computational efficiency.

This design allows the encoder depth to scale naturally with input resolution:
for higher spatial resolutions in $(R,\Theta)$, additional downsampling stages
may be introduced to preserve manageable latent sizes while enabling the
network to represent increasingly complex spatial structure.
\subsection{Details of Temporal TCN}
\subsubsection{TCN Input Sequence}

For each input window ending at time~$t$, the temporal module operates on a
sequence of latent representations,
\begin{equation}
    \mathbf{F}
    =
    \left[ f_{t-T_c+1},\ldots,f_t \right]
    \in \mathbb{R}^{B\times T_c\times d_{\mathrm{lat}}},
\end{equation}
which is provided as input to the temporal convolutional network (TCN).
%This formulation ensures that all predictions depend exclusively on past information,
%consistent with the causal structure of gyrokinetic turbulence evolution.

\subsubsection{Temporal Projection and Dilated Residual Modeling}

The latent sequence is first mapped to an internal temporal feature space via a
learned one-dimensional convolutional projection,
\begin{equation}
    d_{\mathrm{lat}} \;\longrightarrow\; d_{\mathrm{temp}},
\end{equation}
which prepares the features for temporal processing. The projected sequence is
then passed through a stack of residual dilated convolutional blocks, following
standard TCN design principles~\cite{bai2018tcn,oord2016wavenet}.

\subsubsection{\label{app:TCN}Receptive field for a dilated TCN}

In general for multi-layer casually dilated TCN, the dilation factor at layer $\ell$ is chosen according to a
geometrically increasing schedule,
\begin{equation}
    d_\ell = d_0\,\alpha^{\ell}, \qquad \alpha > 1,
\end{equation}
which enables the receptive field to grow exponentially with network depth while
preserving computational efficiency. For a convolutional kernel of size $k$, a
single dilated convolution applied at index $t$ may be written as
\begin{equation}
    y_t
    =
    \sum_{j=0}^{k-1}
        w_j\, x_{t - j d_\ell},
\end{equation}
with an effective receptive field
\begin{equation}
    R_{\mathrm{eff}}^{(\ell)} = (k-1)\, d_\ell + 1.
\end{equation}
Stacking multiple dilated layers results in a total receptive field that grows
rapidly with depth, allowing the TCN to capture multi-scale temporal dependencies
across discrete snapshots without requiring long sequential models.

\subsubsection{Causality and Temporal Latent Representation}

Causality is enforced through left-padded convolutions that prevent access to
future inputs. The TCN produces a sequence of hidden states,
\begin{equation}
    h_{1:T_c} = \mathrm{TCN}(\mathbf{F}),
\end{equation}
from which a single temporally aggregated latent representation is extracted,
\begin{equation}
    z_T = \mathcal{A}\!\left(h_{1:T_c}\right),
\end{equation}
where $\mathcal{A}(\cdot)$ denotes an aggregation operator. In practice,
this representation encodes short-range temporal dependencies relevant for future field and transport prediction while remaining agnostic to long-term phase evolution.
\subsubsection{Temporal TCN for Latent Evolution}

The spatial encoder produces a latent vector $f_t \in \mathbb{R}^{d_{\mathrm{lat}}}$ for each
time slice of the turbulent field.  
To incorporate temporal dynamics, these latent vectors are processed by a 
temporal convolutional network (TCN), a class of 1D architectures that employ dilated convolutions to model long-range dependencies with stable
gradient behavior \cite{bai2018tcn,oord2016wavenet,yu2016dilated,lea2017temporal}.  
TCNs provide an attractive alternative to recurrent models because they support
parallel sequence computation and controllable receptive fields.

\subsection{Two-Head Implementation for Flux and Field Prediction}

The latent state $z_T$ is shared by two prediction heads:

\begin{enumerate}
    \item \textbf{Flux head:} a multilayer perceptron producing the multispecies
    turbulent transport
   \begin{equation}
     z_T \mapsto 
    \widehat{\mathbf{F}} = 
    \left[ (Q_i, Q_e, \Gamma)_{t+1}, \ldots, (Q_i, Q_e, \Gamma)_{t+H} \right] \notag
\end{equation}
over one or multiple prediction horizons $H$. This yields a tensor of shape $[B,H,3]$.  
The flux head depends only on the TCN-evolved latent, providing a low-cost mapping compared to 
full gyrokinetic flux computations.

    \item \textbf{Field-evolution head:} a lightweight convolutional decoder
    predicting the coarse future electrostatic potential $\widehat\Phi(t+H)$ as part of
    an autoregressive rollout mechanism.
\end{enumerate}

Using a shared TCN ensures that both heads rely on a consistent, causally
structured temporal representation, while guaranteeing that predictions never
incorporate information from times $t'>t$.

\subsection{3D U-Net decoder for predicting $\widehat\Phi(t+H)$}

To reconstruct the future gyrokinetic potential field, a three-dimensional
decoder maps the temporally aggregated latent representation $z_T$ back to
physical space through a U-Net–style upsampling pathway. The latent vector is
first reshaped into a coarse spatial feature map compatible with the deepest
encoder level, after which a sequence of spatial upsampling stages progressively
restore resolution in the $(R,\theta)$ directions.

At each decoder stage, the upsampled features are fused with the corresponding
encoder activations via skip connections, allowing fine-scale phase information
from earlier levels to complement the increasingly global representations carried
by the latent state. This multiscale fusion enables the decoder to reconstruct
both large-scale structure and small-scale turbulent features in the predicted
field.

The final decoder layer produces a complex-valued prediction of the gyrokinetic
potential at a future time offset $t+H$, yielding $\widehat{\Phi}(t+H)$ with real
and imaginary components defined on the original spatial grid.

\vspace{1ex}
\paragraph{Upsampling strategy.}
Spatial upsampling within the decoder is performed using deterministic
interpolation followed by convolutional refinement, rather than learned
transposed convolutions. This choice is motivated by both numerical stability and
physical fidelity. Transposed convolutions are known to introduce grid-dependent
artifacts arising from uneven kernel overlap, which can manifest as spurious
small-scale structure and unphysical high-wavenumber content in reconstructed
fields. In contrast, interpolation-based upsampling provides a smooth and
well-conditioned increase in spatial resolution, ensuring consistent alignment
with encoder features prior to skip-connection fusion. This approach is particularly well suited to gyrokinetic turbulence, where upsampling is required only in selected spatial dimensions while spectral modes in other directions must be preserved without modification. 

\section{\label{app:coh}Phase–Coherence Diagnostic}

To quantify the similarity between the predicted electrostatic potential 
$\hat{\phi}(\mathbf{r},t)$ produced by the surrogate model and the 
ground–truth CGYRO potential $\phi(\mathbf{r},t)$, we employ two 
complementary diagnostics widely used in turbulence analysis: 
(i) a $k_y$--resolved phase–coherence measure, and 
(ii) a comparison of the angle–averaged fluctuation spectrum 
$\langle |\phi|^2 \rangle_{r,\theta}(k_y)$ (see Fig.~\ref{fig:dist}). In all expressions below we treat $\phi$ as a complex field,
$\phi = \phi_{\mathrm{real}} + i\,\phi_{\mathrm{imag}}$,
constructed directly from CGYRO’s real and imaginary Fourier components.

For each toroidal mode $k_y$, we compute the local phase difference
between prediction and truth:
\begin{equation}
    \Delta\varphi(\mathbf{x},k_y)
    = 
    \arg\!\left[
        \hat{\phi}(\mathbf{x},k_y)\,
        \phi^*(\mathbf{x},k_y)
    \right],
\end{equation}
where $\mathbf{x}\equiv (r,\theta)$ denotes the remaining spatial indices.
A perfectly phase–aligned prediction yields 
$\Delta\varphi = 0$ everywhere, whereas phase–random errors lead to 
a uniform distribution over $[-\pi,\pi]$.

A robust scalar measure of phase similarity is the 
\emph{circular coherence}, defined as the magnitude of the complex average
over space and ensemble realizations:
\begin{equation}
    C_\varphi(k_y)
    =
    \left|\,
        \left\langle
            e^{\,i\,\Delta\varphi(\mathbf{x},k_y)}
        \right\rangle_{\mathbf{x},\,\text{ens}}
    \right|.
\end{equation}
This quantity satisfies $0 \le C_\varphi \le 1$.
Values $C_\varphi \approx 1$ indicate that the surrogate model reproduces
the correct turbulent phase structure at that $k_y$, a particularly
stringent requirement for gyrokinetic turbulence where radially--extended
ballooning eigenstructures and nonlinear phase alignment drive transport
\cite{barnes2010,waltz1994}.

\section{Loss functions}
\label{app:loss}
The training objective consists of multiple complementary terms designed to
balance instantaneous accuracy and short-horizon stability.

\paragraph{Flux prediction loss.}
For each forecast horizon $\tau_h$, a mean-squared error (MSE) loss is applied to
the predicted transport fluxes,
\[
\mathcal{L}_{\mathrm{flux}}
= \frac{1}{H} \sum_{h=1}^{H}
\left\| \widehat{\mathbf{Q}}(t+\tau_h) - \mathbf{Q}(t+\tau_h) \right\|_2^2 ,
\]
where $\mathbf{Q} = (Q_i, Q_e, \Gamma)$ denotes the target flux vector.

\paragraph{Field reconstruction loss.}
To encourage accurate short-term field evolution, an MSE loss is applied to the
predicted gyrokinetic potential at the next time step,
\[
\mathcal{L}_{\Phi}
= \left\| \widehat{\Phi}(t+1) - \Phi(t+1) \right\|_2^2 .
\]

\paragraph{Rollout consistency loss.}

An optional auxiliary loss is introduced to promote stability under autonomous
multi-step prediction. Over a rollout window of length $K$, this term penalizes
deviations between predicted and reference quantities,
\begin{align}
\mathcal{L}_{\mathrm{roll}}
=
\lambda_{\mathrm{roll}}\Bigg[
&\frac{1}{K}\sum_{k=1}^{K}
\left\|
\widehat{\mathbf{Q}}(t+k) - \mathbf{Q}(t+k)
\right\|_2^2 \nonumber \\
&+ \alpha_{\Phi}\,
\frac{1}{K}\sum_{k=1}^{K}
\left\|
\widehat{\Phi}(t+k) - \Phi(t+k)
\right\|_2^2
\Bigg],
\end{align}
where predictions are evaluated recursively over the rollout horizon. This term
improves autoregressive robustness while preserving single-step accuracy.

\paragraph{Combined objective.}
The total loss minimized during training is
\[
\mathcal{L}
= \mathcal{L}_{\mathrm{flux}}
+ \alpha_{\Phi}\,\mathcal{L}_{\Phi}
+ \mathcal{L}_{\mathrm{roll}},
\]
with the relative weights $\alpha_{\Phi}$ and $\lambda_{\mathrm{roll}}$ selected
on a per-experiment basis.

%\section*{Acknowledgments}
%This work used resources of NERSC (DOE Office of Science). Conducted independently;
%no institutional endorsement is implied.

\section*{References}
\bibliography{CGYRONN}
\end{document}